\documentclass[12pt]{article}
\usepackage{amssymb,amsmath,amsthm,mathrsfs,latexsym,amsxtra,graphicx,appendix,hyperref,setspace,subfig,fix-cm,color,cite,dsfont,longtable,multirow,tikz,longtable,array,fnpct}
\numberwithin{equation}{section}

\newcommand{\bea}{\begin{eqnarray}}
\newcommand{\eea}{\end{eqnarray}}
\newcommand{\be}{\begin{equation}}
\newcommand{\ee}{\end{equation}}

\newcommand{\nn}{\nonumber}

\newcommand{\cF}{\mathcal F}
\newcommand{\cz}{\mathcal Z}
\newcommand{\bz}{\mathbb Z}

\newcommand{\QQ}{{\mathbb{Q}}}
\newcommand{\bt}{\mathbb T}
\newcommand{\ch}{\mathcal H}
\newcommand{\uno}{{\mathds 1}}

\newcommand{\tdl}{{\mathrm{\Gamma}(16+2d,2d)}}
\newcommand{\tDl}{{\mathrm{\Gamma}(16+D,D)}}
\newcommand{\op}{\hspace{1pt}}
\newcommand{\opm}{\hspace{-3pt}}

\newcommand{\sbig}{\mbox{\large$s$}}
\newcommand{\vs}[1]{\vspace{#1pt}}

\newcommand{\msm}[1]{\mbox{\small$#1$}}

\newcommand{\vtgen}[2]{\vartheta\!
\Big[\!
	\hspace*{-1mm}\begin{array}{c}
		\msm{{#1}}\\[-1mm]
		\msm{{#2}}
	\end{array}
	\hspace*{-1mm}\!\Big]}
\newcommand{\hvtgen}[2]{\widehat\vartheta
\Big[\!
	\hspace*{-1mm}\begin{array}{c}
		\msm{{#1}}\\
		\msm{{#2}}
	\end{array}
	\hspace*{-1mm}\!\Big]}

\begin{document} 

\begin{titlepage}
\begin{center}

\hfill \\
\hfill \\
\vskip 0.75in

{\Large 
	\bf Asymmetric Orbifolds, Rank Reduction and\\[8pt] Heterotic Islands
}\\

\vskip 0.4in

{\large Gerardo Aldazabal${}^{a,b,c}$, Eduardo Andr\'es${}^{a,c}$,\\[2mm] Anamar\'ia Font${}^{d,e}$, Kumar Narain${}^{f}$, and Ida G.~Zadeh${}^{g}$
}\\
\vskip 4mm

${}^{a}$
{\it G. F\'isica CAB-CNEA, Centro At\'omico Bariloche, Av. Bustillo 9500, Bariloche, Argentina} \vskip 1mm
${}^{b}$
{\it Consejo Nacional de Investigaciones Cient\'ificas y T\'ecnicas {\rm(}CONICET{\rm)}} \vskip 1mm
${}^{c}$
{\it Instituto Balseiro, Universidad Nacional de Cuyo {\rm(}UNCUYO{\rm)}, Av. Bustillo 9500, R8402AGP,  Bariloche, Argentina} \vskip 1mm
${}^{d}$
{\it Fac. de Ciencias, Universidad Central de Venezuela, A.P.20513, Caracas 1020-A, Venezuela} \vskip 1mm
${}^{e}$
{\it Max-Planck-Institut f\"ur Gravitationsphysik, Albert-Einstein-Institut, 14476 Golm, Germany} \vskip 1mm
${}^{f}$
{\it International Centre for Theoretical Physics, Strada Costiera 11, 34151 Trieste, Italy} \vskip 1mm
${}^{g}$
{\it Mathematical Sciences \& STAG Research Centre, University of Southampton, Southampton SO17 1BJ, UK} \vskip 1mm

\end{center}

\vskip 0.35in

\begin{center} {\bf ABSTRACT } \end{center}

We consider toroidal asymmetric orbifolds of the heterotic string preserving all 16 supercharges, developing
a general formalism to study components of the moduli space characterized by rank reduction of the gauge
group. In particular we construct six- and four-dimensional heterotic islands with no massless moduli other than the dilaton.
The formalism involves the Leech lattice, its automorphisms and their corresponding invariant and normal, or coinvariant, 
sublattices.

\vfill

\noindent{January 28, 2025}

\end{titlepage}

\setcounter{page}{1}
\setcounter{tocdepth}{2}

\tableofcontents

\section{Introduction}\label{section_intro}

Exploring unknown regions of the moduli space of world-sheet conformal field theories is an important part of collective efforts to 
understand quantum gravity at a fundamental level. Even when supersymmetry is fully preserved and powerful tools are available, there 
are loci in the moduli space which have remained hidden from the light of traditional methods. Some of these loci include heterotic vacua 
with reduced rank of the gauge group, whose world-sheet descriptions are not yet known. The extreme case of such vacua are called 
`islands' because they have no massless vector multiplets, and therefore no continuous moduli except for the dilaton \cite{Dabholkar:1998kv}. 
The goal of this work is to formulate a world-sheet description of fully supersymmetric compactifications of heterotic strings with 
rank reduction, including islands, within the framework of toroidal asymmetric orbifolds \cite{Narain:1986qm, Narain:1990mw}. 
To this end we will build on our approach developed earlier in heterotic non-supersymmetric setups \cite{Acharya:2022shu}. 
Although the emphasis will be on heterotic compactifications with 16 supercharges, our formalism can be applied to compactifications 
with less supersymmetry, and to type II strings.

String vacua with 16 supercharges and reduced rank were originally discovered in type I toroidal compactifications \cite{Bianchi:1991eu}.
In the heterotic case such vacua were first obtained in the fermionic formulation and came to be known as CHL 
strings \cite{Chaudhuri:1995fk}. The moduli space component with rank reduced by eight was later realized in the 
bosonic formulation in terms of $\bt^D/\bz_2$ asymmetric orbifolds in \cite{Chaudhuri:1995bf} and analyzed in more detail 
in \cite{Mikhailov:1998si}, see also \cite{deBoer:2001wca, Font:2021uyw, Fraiman:2021soq, Fraiman:2021hma, Fraiman:2022aik}.
Generalizations to other $\bt^D/\bz_n$ asymmetric orbifolds have been considered e.g. in 
\cite{Chaudhuri:1995dj, deBoer:2001wca, Persson:2015jka, Bossard:2017wum, Fraiman:2021soq}. 
Vacua with 16 supercharges and rank reduction can 
also be devised as type II compactifications on toroidal asymmetric orbifolds. Examples include rank 1 theories in 9 dimensions
\cite{Bianchi:1997rf, Aharony:2007du} and the rank-0 Dabholkar-Harvey island in 6 dimensions \cite{Dabholkar:1998kv}.
Further interesting progress on asymmetric orbifold constructions in string theory in the last years may be found in 
\cite{Aoki:2004sm, Beye:2013moa, Tan:2015nja, Satoh:2015nlc, GrootNibbelink:2017usl, Harvey:2017rko, Harvey:2017xdt, 
GrootNibbelink:2020dib, Baykara:2021ger,  Bianchi:2022tbr, Cheng:2022nso, Montero:2022vva, Gkountoumis:2023fym, Nakajima:2023zsh, 
Faraggi:2023mkn, Baykara:2023plc,  Israel:2023tjw, Gkountoumis:2024dwc, Gkountoumis:2024boj, Baykara:2024vss, 
Baykara:2024tjr, Angelantonj:2024jtu}. For relevant earlier studies, see \cite{Ibanez:1987pj, Lust:1987ts, Harvey:1987da}.

An important feature of supergravity theories with 16 supercharges is that the rank of the gauge group is restricted by 
quantum consistency conditions such as anomaly cancellation and unitarity of the theory on string or brane probes. Notably,
in 10 dimensions it has been shown that only rank 16 is allowed \cite{Adams:2010zy, Kim:2019vuc}. Furthermore, in $(10-D)$ 
dimensions the upper bound $(16+D)$ was established in \cite{Kim:2019ths}, and when $D=1,2$, it has been argued that only 
ranks $(16+D)$, $(8+D)$ and $D$ can occur \cite{Montero:2020icj, Bedroya:2021fbu, Lao:2023tuj}. 
In this work we do not directly address the question of allowed ranks, except for remarking that our results align 
with the empirical observation that in $(10-D)$ dimensions, with $D > 2$, the rank is even (odd) when $D$ is even (odd). 

This paper rather aims at populating the landscape of heterotic string compactifications with 16 supercharges by providing explicit 
models with different ranks in diverse dimensions. In particular, we will present $\bt^4/\bz_{2m}$ asymmetric orbifolds, 
with $m=1,3,5$, and rank 8,2,0 respectively. The existence of such orbifolds in the moduli space of six-dimensional theories with 
16 supercharges was conjectured in \cite{Fraiman:2022aik}, but to the best of our knowledge, their string theory realization has been
lacking until now. The main outcome of this work will be a general formalism for constructing heterotic compactifications with 
16 supercharges and rank reduction in terms of $\bt^D/\bz_n$ asymmetric orbifolds. The specific $6d$ orbifolds mentioned above, 
including the island case of rank-0, will be concrete examples within this approach. We will also describe a group or `archipelago' of 
heterotic islands in four dimensions, emerging from a $\bt^6/\bz_{22}$ orbifold.

An essential component of heterotic asymmetric orbifolds $\bt^D/\bz_n$ is the even self-dual lattice $\Gamma$ with signature 
$(16+D,D)$, which encodes the values of the left- and right-moving momenta of the compact bosons.
The generator of the orbifold group $\bz_n$ acts on these momenta, leaving invariant a sublattice of $\Gamma$ that we denote by $I$.
Consequently, vectors in the orthogonal complement of $I$ in $\Gamma$, called normal or coinvariant lattice and
denoted $N$, are rotated under the action of $\bz_n$. Since we assume that all supersymmetries are unbroken this
normal lattice will have definite signature, concretely purely left-moving in our conventions.
From the formalism of asymmetric orbifolds we can already draw several important conclusions about the resulting theory 
in $(10-D)$ dimensions. To begin we notice that the spectrum in the orbifold untwisted sector will not contain massless 
gauge multiplets from bosonic oscillator modes along the rotated $N$ directions. However, this does not imply that the rank of 
the gauge group will be reduced, as massless gauge multiplets can still arise from invariant combinations of roots in $N$.
Hence, a necessary condition for rank reduction is that $N$ does not have roots, i.e. vectors with length squared equal to 2. 
Another immediate upshot is that for rank-0 islands without gauge multiplets, the invariant lattice must have definite signature 
$(0,D)$, since invariant left-moving directions would give rise to massless gauge multiplets from oscillator excitations. Therefore, 
for rank-0 both $I$ and $N$ are rigid, i.e. they do not have geometric moduli as expected for an island.
These are actually necessary but not sufficient conditions for rank reduction. We also have to study whether extra conditions 
must be satisfied to remove massless states in the twisted sectors.

The result that rank reduction requires absence of root vectors in the normal lattice motivates us to base our analysis on sublattices of 
the Leech lattice, which is the unique definite even self-dual lattice of rank 24 without any roots. 
The Leech lattice has a large group of automorphisms, the Conway group $\text{Co}_0$.
The classification of invariant sublattices under the action of $\text{Co}_0$ done by H\"ohn and Mason \cite{HM} will play
a central role in our approach to constructing heterotic asymmetric orbifolds with rank reduction.
The Leech lattice has appeared previously in various contexts within string theory and conformal field theory. 
Specifically, reference \cite{Chaudhuri:1995ee} discussed $2d$ heterotic models involving cyclic orbifolds of the Leech lattice 
and proposed decompactification to higher dimensions.
In \cite{Gaberdiel:2011fg}, the Leech lattice was instrumental in the classification of supersymmetry-preserving symmetries of 
non-linear sigma-models on K3. The authors of \cite{Harvey:2017xdt, Baykara:2021ger} utilized sublattices of the Leech lattice to study 
discrete symmetries and dualities of heterotic strings compactified on tori and orbifolds. 
In this work we will construct reduced-rank theories in $(10-D)$ spacetime dimensions by
implementing automorphisms and sublattices of the Leech lattice into $\bt^D/\bz_n$ asymmetric orbifolds of the heterotic string.

Once the momentum lattice $\Gamma$ and the $\bz_{n}$ automorphism are set up from Leech lattice data, we 
will compute the partition function of the asymmetric orbifold. The untwisted sector terms depend on the action
of the orbifold generator on the original $\bt^D$, which typically involves a translation in $\Gamma$ by 
a constant shift along the invariant directions. The twisted sector terms are derived by applying modular transformations.
When the order $n$ is odd, modular invariance just gives a condition on the constant shift. However, when $n$ is even there
are additional constraints as stated in \cite{Narain:1986qm, Narain:1990mw}, and discussed more recently in 
\cite{Harvey:2017rko,Taormina:2017jnt}. We will specialize to the case $n=2m$, where $m$ is an odd prime number, 
which is of interest and affords a general treatment. We will find that modular invariance can be achieved when the orbifold
action on $\Gamma$ takes into account a vector that characterizes the $\bz_2$ conjugacy classes of $I^*/I$
with half-integer squared norm.

The plan of this paper is as follows. In section \ref{section_asym} we describe the class of asymmetric orbifolds that enable rank 
reduction. We first explain how the underlying even self-dual lattice $\Gamma$ and the automorphism defining the orbifold
are determined from Leech lattice data. We then deduce key properties of the sublattices of $\Gamma$ that are left invariant by 
$\bz_{2m}$ automorphisms and their powers.
In section \ref{section_Z} we explicitly compute the partition function of heterotic $\bt^D/\bz_{2m}$ asymmetric orbifolds.
To streamline the analysis we focus on the case in which the invariant lattice under the
$\bz_{2m}$ automorphism has definite signature, which is actually a necessary condition for islands in our setup.
The case of invariant lattice with indefinite signature is left for an appendix. 
We will first discuss the contribution of the untwisted sector to the partition function and then apply modular transformations to
obtain the twisted sector terms.
The constraints implied by modular invariance in each sector are derived and summarised in subsection \ref{subsec_sum}. 

In section \ref{section_op} we examine the operator interpretation of the partition function in all twisted sectors. We obtain
the complete set of consistency conditions and confirm that they are satisfied, thereby ensuring the positive integer 
multiplicity of states. Additionally, we verify that the untwisted sector partition function also entails positive integer multiplicities.
Section \ref{section_hm} presents specific examples of heterotic compactifications with reduced rank, including islands. 
In subsection \ref{section_hm149} we discuss two islands in 6 dimensions, constructed as $\bt^4/\bz_{10}$ orbifolds with
different invariant lattices. In subsection 
\ref{section_hm251} we consider a $\bt^6/\bz_{22}$ orbifold leading to a $4d$ archipelago. Subsection \ref{section_hm100} 
reports a $\bt^4/\bz_{10}$ orbifold giving rise to a $6d$ theory with reduced rank equal to four. 
We close in section \ref{section_final} by outlining our findings and reflecting on future directions.

There are several appendices accompanying the main text. 
In appendix \ref{app_w} we give a constructive proof of the existence of the characteristic vector of $I^*/I$. 
Appendices \ref{app_g2}-\ref{app_Lsum} contain intermediate results, as well as proofs of various lattice properties 
and identities, needed in sections \ref{section_Z} and \ref{section_op}. In appendix \ref{appnoneqv} we  comment on the 
requirements on constant shift vectors to lead to non-equivalent models.
In appendix \ref{apprr} we discuss $\bt^4/\bz_{2m}$ orbifolds, with $m=1,3$, in which the invariant lattice is indefinite
and the rank is reduced as proposed in \cite{Fraiman:2022aik}.

\section{A class of heterotic asymmetric orbifolds}\label{section_asym}

In this section we first briefly review the basics of asymmetric orbifolds \cite{Narain:1986qm, Narain:1990mw},
introducing conventions and notation along the way. We will then describe a class of models with unbroken
supersymmetry and reduced rank of the gauge group. 
The main elements needed to construct explicit examples will be presented in detail.

We start with the 10-dimensional supersymmetric heterotic string. 
For the left-moving and right-moving sectors we respectively take the bosonic string 
and the superstring degrees of freedom. The right-moving world-sheet fermions will be
described in terms of $SO(8)$ bosons, with the GSO projection selecting weights in
the vector ($V$) and the spinor ($Sp$) classes. 

In the next step we compactify the theory on $\bt^{D}$ and mod out by an action that treats left and right movers 
differently. 
According to the formalism of asymmetric orbifolds \cite{Narain:1986qm, Narain:1990mw},
we need to define the action of the orbifold generator on the left- and right-moving momenta taking values in the even 
self-dual lattice $\tDl \equiv \Gamma$, as we now discuss.

The action on $\Gamma$ is given by some automorphism $\Theta$ of order $n$ which does not mix left and right movers,
i.e. \mbox{$\Theta=(\Theta_L, \Theta_R)$}. In general $\Theta$ only exists at a slice in the moduli space
of $\Gamma$, parametrized by the background values of the $\bt^{D}$ metric, the Kalb-Ramond $B$ field and the
Wilson lines. In order to describe its action, it is convenient to introduce
the sublattice $I$ left invariant by $\Theta$, i.e.
\be\label{I}
I=\big\{P \in \Gamma \, ||\,  \Theta P = P \big\} \ .
\ee
We also introduce the normal sublattice $N$, defined as the orthogonal complement of $I$ in $\Gamma$, namely
\be
N=\big\{P \in \Gamma \, || \, P\cdot X = 0,  \, \forall\op X \in I \big\}\ .
\ee
$N$ is also known as the coinvariant lattice. The dual lattices of $I$ and $N$ are denoted $I^*$ and $N^*$.
The invariant and the normal lattice are even.

Any vector $P \in \Gamma$ may be
written as $P\!=\!(P_N, P_I)$, with $P_N \!\in\! N^*$ and $P_I \!\in\! I^*$ \cite{Narain:1986qm}. 
Furthermore, using that $\Gamma$ is even self-dual it can be shown that
\be
N^*/N \cong I^*/I \ ,\qquad q_I\circ \varsigma=-q_N\ ,
\ee
where $q_L$ is the discriminant 
form\footnote{For an even lattice $L$, the discriminant form is a map $q_L: L^*\!/L \to \QQ/2\bz$, $x+ L \mapsto x^2\, \text{mod}\, 2$.} 
of lattice $L$,
and $\varsigma$ is a one to one pairing of the cosets.
This means that
for a coset representative $\mathrm{w} \in N^*\!/N$ there exists $\varsigma(\mathrm{w})
\in I^*\!/I$ such that
\mbox{$q_I(\varsigma(\mathrm{w}))=-q_N(\mathrm{w})$}.
The lattice $\Gamma$ itself
can be constructed adding the correlated classes \cite{Nikulin80int}. Schematically,
\be\label{Gglue}
\Gamma = (N, I) + \coprod_{\mathrm{w} \in N^*\!\!/N} \hspace*{-2mm} (\mathrm{w}, \varsigma(\mathrm{w}))\ .
\ee
The glue vectors $(\mathrm{w}, \varsigma(\mathrm{w}))$ have even norm and integer scalar product with each other. A more detailed expression for $\Gamma$ is given in \eqref{Gamma}.

The decomposition of $\Gamma$ into the direct sum of normal and invariant lattices plus correlated classes 
will be a key ingredient in our construction of asymmetric orbifolds. One important reason is that it encodes
the momenta of untwisted and twisted string states. Hence it has implications for the modular invariance of the
partition function. Moreover, as we will explain shortly, it enters in the prescription to choose the automorphism $\Theta$.

Let us now specialize to heterotic asymmetric orbifolds $\bt^{D}/\bz_n$ preserving all supersymmetries.  
In this case the $\bz_{n}$ automorphism of $\tDl$ does not act on the $D$ right movers but only on 
some number $s$ of left movers. This means $\Theta_R=\uno$ and $\Theta_L=\Theta$.
Therefore, the invariant lattice $I$ has signature $(16+D-s,D)$, whereas the normal lattice $N$ has signature 
$(s,0)$, with $0 \leq s \leq 16+D$.

In the purely toroidal compactification, massless fields of the resulting $(10-D)$-dimensional theory arrange into the
supergravity multiplet, which includes $D$ $U(1)$ graviphotons, and \mbox{$(16+D)$} $U(1)$ vector multiplets at generic points
of moduli space. At special points there can be enhancement to some non-Abelian group of rank $(16+D)$.
In the asymmetric orbifolds the rank can be reduced. For instance, at a point in moduli space
such that $\tDl \sim E_8 \oplus E_8 \oplus DU$, where $U$ denotes the even self-dual $(1,1)$ lattice,
we can perform a $\bz_2$ orbifold that acts by exchange of the 2 $E_8$'s
and translation by half a period along one of the circles. This setup gives CHL models with gauge
group of rank $(8+D)$, excluding graviphotons \cite{Chaudhuri:1995bf}.

We are interested in generic features and explicit construction of models with rank reduction.
The formalism of asymmetric orbifolds already leads to the simple observation that a necessary 
requirement for rank reduction is that the normal lattice $N$ does not have roots,  i.e. vectors of length squared 2.
To understand this statement it is useful to recall the structure of massless states in the untwisted sector.
By GSO projection,  the right-moving piece  consists of $\mathbf{8_v}$ and $\mathbf{8_s}$ decomposed under $SO(8-D)$. 
On the other hand, the left-moving piece includes oscillator modes with occupation number $N_L=1$ or momentum
states with $P_L^2=2$. Now, since the $s$ directions along $N$ are not invariant under the action of $\Theta$, the 
corresponding left moving oscillators cannot give massless states belonging to vector multiplets. Thus, naively we would 
conclude that the rank is reduced by $s$. However, if $N$ has root vectors then we could construct invariant combinations 
that give additional massless vectors thereby increasing the rank. 
The condition that $N$ has no roots is necessary but not sufficient for rank reduction of the gauge group because 
massless vectors could arise in twisted sectors. As we will explain, extra massless twisted states might or might not be avoided
by including appropriate constant translations in the lattice which are compatible with modular invariance.

In the CHL models described above, the normal 
lattice is\footnote{$L(j)$ means lattice $L$ with Gram matrix rescaled by a factor $j$.} $E_8(2)$, 
which indeed does not have root vectors. This is also the case in other known examples in 7 dimensions 
constructed as heterotic asymmetric orbifolds $\bt^2 \times S^1/\bz_n$, $n=3,4,5,6$ \cite{deBoer:2001wca, Fraiman:2021soq}. 
The $\bt^2$ is chosen to be rectangular, the $B$ field is set to zero, and two specific Wilson lines are turned on.
The full orbifold action includes an order $n$ translation along the $S^1$, together with an explicit
$\bz_n$ automorphism of $\Gamma(18,2)$. Given this automorphism and the momenta in $\Gamma(18,2)$ we
can find the invariant sublattices and show that the associated normal lattices do not have roots. 
For instance, for $n=3$ the normal lattice turns out to be the 12-dimensional Coxeter-Todd lattice. This result is
known in the Mathematics literature \cite{Garbagnati07, Garbagnati09}.  In fact, these \mbox{7-dimensional} models
can be equivalently realized as heterotic asymmetric orbifolds $\bt^3/\bz_n$, where $\bz_n$ is an
automorphism of the $\Gamma(19,3)$ lattice. Abelian automorphisms of $\Gamma(19,3)$ with invariant lattice
of signature $(19-s,3)$ and normal lattice of signature $(s,0)$ without roots, correspond to symplectic automorphisms of
K3 surfaces which have been classified \cite{Nikulin79}. Besides $n=2, \dots, 6$, such $\bz_n$ automorphisms 
exist for $n=7,8$, but in these cases in the corresponding $\bt^3/\bz_n$ asymmetric orbifolds,
massless states in twisted sectors cannot be avoided. However, for $n=7,8$, it is possible to construct $\bt^3 \times S^1/\bz_n$
asymmetric orbifolds with rank reduction by combining the $\Gamma(19,3)$ automorphism with an order $n$
translation along the circle.

\subsection{Choice of \texorpdfstring{$\Gamma$}{TEXT} and automorphism}\label{subsection_leech}

Our strategy to specify the automorphism $\Theta$ is adapted to our motivation to construct theories with 
rank reduction, in particular string islands without vector multiplets. As already argued, this requires that the normal lattice
does not have vectors of length squared equal to 2. We are then led to consider automorphisms of the Leech
lattice, denoted $\Lambda$, which is the unique \mbox{24-dimensional} even self-dual lattice without root vectors. 
The idea to embed the normal lattice in $\Lambda$ was previously used to study symmetries of K3 sigma models,
which involve the $\Gamma(20,4)$ lattice \cite{Gaberdiel:2011fg}. Heterotic compactifications with rank reduction
based on orbifolds by automorphisms of $\Lambda$ had been explored earlier \cite{Chaudhuri:1995ee}.

The automorphism group of $\Lambda$ is the Conway group $\text{Co}_0$ \cite{Conway98}.
It has been shown by H\"ohn and Mason that there are 290 classes of sublattices of $\Lambda$ that are left fixed
by elements of $\text{Co}_0$ \cite{HM}. In Appendix A of the published version, they have also supplied the 
basis for the corresponding invariant and coinvariant lattices in Magma format \cite{magma}. 
We will refer to the classes as \mbox{HM\#}, where \# is the row number in Table 1 of \cite{HM}.

The invariant and coinvariant lattices listed by H\"ohn and Mason, which we denote $\tilde I$ and $N$, are both Euclidean, 
i.e. they have signatures $(24-s,0)$ and $(s,0)$ respectively. In our language $N$ is the normal lattice that we want to use to 
construct an even self-dual lattice $\Gamma$ of signature $(16+D,D)$. To this end we have to search for another lattice $I$ of 
signature $(16+D-s,D)$ such that $N^*/N \cong I^*/I$, $q_I\circ\varsigma=-q_N$, and  the classes  of $I^*/I$ can be correlated with 
the classes of $N^*/N$. The discriminant group $N^*/N$, as well as a basis of generators, can be obtained
from the Smith decomposition of the Gram matrix of $N$, as summarized e.g. in proposition 3 of \cite{Baykara:2021ger}.
For the candidate invariant lattice $I$ the procedure is similar. 
A complete set of even and mutually integral glue vectors is obtained by pairing combinations of generators iteratively.
In this way we assemble the lattice $\Gamma$ with the structure in \eqref{Gglue}.
In \cite{Harvey:2017xdt, Baykara:2021ger} the authors also applied this 
scheme to construct $\Gamma$, developing a systematic procedure to find $I$ when $s \ge 16$ by first
looking for embeddings of $\tilde I$ in $E_8$ and then taking
for $I$ the orthogonal complement of $\tilde I$ in $E_8$ with reversed signature.
For the particular case $D=4$ and $s=20$, we find that it is also possible to choose the invariant lattice $I$ to be $\tilde I$ 
with reversed signature. 

Let us now explain how the automorphism $\Theta$ is obtained.
For a given pair $(N,I)$ from which $\Gamma$ is constructed by adding correlated classes, we use the package Magma \cite{magma}
to look for automorphisms of $N$ of order $n$, where $\bz_n$ is one of the factors in the discriminant group $N^*/N$. 
We impose the conditions that the automorphism does not have eigenvalues equal to one and that it preserves the classes 
in $N^*/N$. The latter is necessary to keep the correlations of conjugacy classes of $N^*/N$ with those of $I^*/I$
which are invariant. This then gives an automorphism $\Theta$ of order $n$ of the full $\Gamma$ such that $I$ is the invariant 
lattice and $N$ is the normal lattice. 

Conjugacy classes of the automorphism group $\text{Co}_0$ of $\Lambda$ have also been studied in \cite{HL}, which lists the 
invariant lattice and the associated frame shape that encodes the characteristic polynomial of the particular class \cite{Kondo85}. 
The non-trivial eigenvalues computed from the frame shape coincide with those computed from the explicit $\Theta$ found as explained 
above. These eigenvalues determine the orbifold action on the internal bosonic coordinates and will be needed to compute the
partition function. In general they are of the form $e^{\pm 2\pi i t_a}$, $a=1, \ldots, \frac s2$, where $s$ is the dimension of $N$. 
It is convenient to choose the $t_a$ such that $0 < t_a \leq \frac12$, and to package them as the components of a vector 
$t=(t_1,t_2,\ldots,t_{\frac s2})$. 
Actually, since $\Theta$ acts crystallographically on $N$, the eigenvalues must satisfy known conditions \cite{Erler:1992ki}.
For instance, for a $\bz_{\mathfrak{p}}$ action with ${\mathfrak{p}}$ prime
\vs{-5}\be\label{taprime}
t=\frac{1}{\mathfrak{p}}\left(1,2, \ldots, \tfrac{\mathfrak{p}-1}{2}\right)^{n_1} \ ,
\ee
where the positive integer $n_1$ in the exponent indicates the multiplicity and satisfies  $n_1(\mathfrak{p}-1)=s$.
We will be mostly interested in $\bz_{2m}$ automorphisms with $m$ prime different from 2. In this case we have
\vs{-5}\be\label{ta}
t=\frac{1}{2m}\Big( [1,3, \ldots, m-2]^{n_1}, [2,4, \ldots, m-1]^{n_2}, [m]^{n_m} \Big) \ .
\ee
The exponents are non-negative integers that indicate multiplicity and satisfy the relation \mbox{$(n_1+n_2)(m-1) + 2 n_m=s$}. 
Note that the non-trivial eigenvalues of $\Theta^m$ are $-1$. From the eigenvalues we
can easily evaluate the quantities ${\det}^{'}(1-\Theta^k)$, $k=1,\ldots,m$, where ${\det}'$ excludes eigenvalues equal to 1,
which will enter in the partition function.

The lattice $\Gamma$ formed from the pair $(N,I)$ and correlated classes of $N^*/N \cong I^*/I$, together with the $\bz_{n}$ 
automorphism $\Theta$ are the basic ingredients to construct the heterotic asymmetric orbifold $\bt^{D}/\bz_{n}$. 
In particular, the known heterotic models with rank reduction $\bt^3/\bz_n$, for $n=2,\ldots,6$, can be built starting with the normal lattices 
of the HM2, HM4, HM9, HM20 and HM18 respectively. Based on the HM52 we can also construct  a heterotic asymmetric orbifold 
$\bt^4/\bz_7$ with gauge group of rank 2. In this model the normal lattice of signature $(18,0)$ is isomorphic to the
$\Omega_7$ lattice in \cite{Garbagnati07}, whereas the invariant lattice of signature $(2,4)$ can be 
chosen to be $I=U(7) \oplus K(-1) \oplus U$, where $K$ is a 2-dimensional lattice with Gram matrix 
$\big(\begin{smallmatrix}4& 1 \\1 & 2\end{smallmatrix}\big)$.
The component $U$ in $I$ is the lattice corresponding to a circle and the orbifold can be understood as 
$\bt^3 \times S^1/\bz_7$, as mentioned before.

Other examples with rank reduction will be presented in section \ref{section_hm} and Appendix \ref{apprr} after we discuss 
in more detail the orbifold partition function.

\subsection{Invariant lattices and modular invariance}\label{subsection_I}

In the standard asymmetric orbifold construction the orbifold group has a generator $g$ defined to act on 
$P=(P_N,P_I) \in \Gamma$ as $g|P_N, P_I \rangle = e^{2\pi i P \cdot v} |\Theta P_N, P_I \rangle$, 
where $v$ is a constant shift vector that can be taken along the $I$ directions without loss of generality.
This is actually the action of $g$ in the known untwisted Hilbert space $\ch_0$. The twisted Hilbert space 
$\ch_1$, where strings close up to the action of $g$, is deduced by applying modular transformations.
To explain how this works, let us consider the partition function in the untwisted sector with the insertion of $g$,
i.e.  $\text{Tr}_{\ch_0}\!\! \left(g \, q^{L_0} \, \bar q^{\bar L_0}\right)$. The action of $g$ on $\Gamma$
implies that only momenta with $P_N=0$ survive the trace. In other words, the contribution of momentum
states to the trace will only include a sum over the invariant lattice. Schematically,
\be\label{sch1g}
\text{Tr}_{\ch_0}\!\!
\left(g \, q^{L_0} \, \bar q^{\bar L_0}\right)  \supset
\sum_{P \in I} \hspace*{-.5mm} q^{\frac{1}{2}P_L^2}\, \bar{q}^{\frac{1}{2}P_R^2}\, 
e^{2i\pi P \cdot v} \, .
\ee
We can now do a modular transformation $\tau \to -1/\tau$, which changes the boundary conditions on the
world-sheet, to obtain  $\text{Tr}_{\ch_1}\!\! \left(q^{L_0} \, \bar q^{\bar L_0}\right)$. Poisson resummation leads to
\be\label{schg1}
\text{Tr}_{\ch_1}\!\!
\left(q^{L_0} \, \bar q^{\bar L_0}\right)  \supset
\sum_{P \in I^*} \hspace*{-.5mm} q^{\frac{1}{2}(P+v)_L^2}\, \bar{q}^{\frac{1}{2}(P+v)_R^2}\, .
\ee
This shows that strings states in the $g$-twisted sector will have momenta lying in $I^* + v$.

From \eqref{schg1} we can also infer necessary conditions for modular invariance because if $g$ has order $n$, the 
modular transformation $\tau \to \tau +n$ must leave the trace invariant. This will put restrictions on the shift $v$
that will be spelled out shortly. Now we are more interested in additional requirements on $P \in I^*$ because
the transformation also gives a phase $e^{i \pi n P^2}$ inside the sum. Below we will show that when $n$ is a 
an odd number, $nP^2$ is always even for $P \in I^*$ so that the phase is equal to one. 
However, for $n$ even it is not guaranteed. In this work we will study the case $n=2m$, with $m$ prime, different from 2.
We will see that it is possible to achieve modular invariance with $g$ of order $2m$.

We will compute the full partition function of the theory in section \ref{section_Z}.
In the rest of this subsection we will study properties of $I^*$ in more detail, focusing on the 
case $\Theta^{2m}=\uno$ for $m$ prime, different from 2. In this setup we also need to analyze the invariant lattices under 
$\Theta^2$ and $\Theta^m$, denoted $I_2$ and $I_m$ respectively, and their corresponding dual lattices.

\subsubsection{Conjugacy classes in the dual invariant lattices}

To begin consider $\Theta^n=\uno$ and apply $(1+\Theta+..+\Theta^{n-1})$ on a generic element 
$(P_N,P_I)\in \Gamma$, where $P_N\in N^*$ and $P_I\in I^*$. 
It results in $(0, nP_I) \in I$. Thus, $nP_I \in I$ for all $P_I\in I^*$. This in turn implies that $nP_I \cdot P_I \in \bz$ 
and $(nP_I)^2\in 2\bz$ for all $P_I \in I^*$. Hence, if $n$ is odd it must be $nP_I^2 \in 2\bz$ as claimed before.

Let us now set $n=2m$ with $m$ prime, different from 2. We again obtain that $2m P_I \in I$ for all $P_I$ in $I^*$, but we
cannot conclude that and $2m P_I^2$ is even. Nonetheless we can still extract some useful information
because now we can write $P_I\in\frac1{2m}I=(\frac a2+\frac bm)I$, since there exist pairs of coprime 
integers $a$ and $b$ such that $ma + 2b = 1$. The upshot is that $I^*/I$ generically contains $\bz_2$, $\bz_m$ and 
$\bz_{2m}$ factors, i.e. 
\be\label{I*}
I^*/I\cong N^*/N\cong\bz_2^{\ell'_2}\times\bz_m^{\ell'_m}\times\bz_{2m}^{\ell'_{2m}} 
\cong\bz_2^{\ell_2}\times\bz_m^{\ell_m} \ ,
\ee
where we used that $\bz_{2m} \cong \bz_2 \times \bz_m$. It will be convenient to work with
$\bz_2$ and $\bz_m$ generators, whose numbers are respectively the non-negative integers $\ell_2$ and $\ell_m$. 
We  observe that the  invariant lattice for $\Theta^k$, with $(k,2m)=1$, is also $I$.
 
We can now state a result that will be central to our construction of $\bz_{2m}$ asymmetric orbifolds. 
From the previous arguments it follows that $P_I\in I^*$ can be written as $P_I=X+Y$, for some $X\in\bz_2$ classes and $Y\in\bz_m$ 
classes of $I^*/I$. The vectors $X$ clearly satisfy $2X^2\in\bz$. Although less obvious, it can also be shown that there always exists a 
constant vector $w$ in $\bz_2$ classes of $I^*/I$ such that
\be\label{w}
X^2=X\cdot w\quad \forall\; X\in\bz_2\;\;{\rm classes~of}\;\; I^*/I\ .\vs{-5}
\ee
A constructive proof of existence of $w$ is given in appendix \ref{app_w}. It can further be shown that
\be\label{p2wcond}
e^{2 \pi im P_I^2} = e^{2 \pi iP_I \cdot w} \quad\forall\; P_I\in I^*\ .
\ee
To prove this, notice that $X\cdot Y \in \bz$ because $2X\cdot Y \in \bz$, $X\cdot mY \in \bz$, and $m$ is prime.
Similarly, $m Y^2 \in 2\bz$ because $Y\cdot mY \in \bz$ and $m^2 Y^2 \in 2\bz$. Thus, 
$mP_I^2 = m X^2 \, \text{mod}\, 2$. To arrive at \eqref{p2wcond} we use 
$4X^2 \in 2\bz$, $Y\cdot w \in \bz$, and eq. \eqref{w}. The characteristic vector $W_g$ introduced in \cite{Harvey:2017rko}
has analogous properties as our $w$, being related by $W_g = 2 w \, \text{mod}\, 2$.
The vector $w$ was included in the $\bt^3/\bz_2$ orbifolds constructed in \cite{Acharya:2022shu} where it was shown
that Milgram's theorem \cite[appendix 4]{Milnor73} relates $w^2$ to lattice data as we will also discuss in appendix \ref{appC2}.

Let us now turn to the invariant sublattices $I_m$ and $I_2$ under $\Theta^m$ and $\Theta^2$ respectively. 
Clearly $I$ is contained in $I_m$ and $I_2$. Consider first $I_m$. Any vector in the even self-dual lattice $\Gamma$ can be
expressed as $(P_{I_m^\perp}, P_{I_m})$, where ${I_m^\perp}$ is the orthogonal complement of $I_m$ in $\Gamma$.  
Moreover, $\Gamma$ projected along the directions of $I_m$ and $I_m^\perp$ is respectively $I_m^*$ and $I_m^{\perp*}$. 
Now, acting with $(1+\Theta^m)$ on $(P_{I_m^\perp}, P_{I_m})$ gives $(0,2 P_{I_m}) \in I_m$, which means that only 
$\bz_2$ classes appear in $I_m^*/I_m$. Since the $\bz_m$ classes that were in $I^*/I$ have disappeared, it must be
that {\it all} the $\bz_m$ classes in $N^*/N$ have some representatives that are invariant under $\Theta^m$, so that they 
correlate with $\bz_m$ classes of $I^*/I$ and become part of $I_m$. For this to happen, necessarily the dimension of $I_m$ must be 
bigger than that of $I$ (in other words, some of the directions in $N$ must be invariant under $\Theta^m$).

Similarly, for $I_2$ one may apply  $(1+\Theta^2+\Theta^4+\ldots+\Theta^{2(m-1)})$ on $(P_{I_2^\perp}, P_{I_2})$. 
The result is that $I_2^*/I_2$ contains only $\bz_m$ classes and the $\bz_2$ classes that appeared in $I^*/I$ are correlated with 
some representatives of $\bz_2$ classes appearing in $N^*/N$ that are invariant under $\Theta^2$.

From the above discussion we have learned that the we can choose generators of the $\bz_2$ classes of $N^*$ 
that are $\Theta^2$-invariant. Let us denote them by $f^N_{2,i}$, $i=\{1,\ldots,\ell_2\}$. 
Likewise, we can pick  generators of $\bz_m$ classes of $N^*$ that are $\Theta^m$-invariant and denote them by 
$f^N_{m,j}$, $j=\{1,\ldots,\ell_m\}$. These generators satisfy 
\be\label{invV2Vm}
\Theta^2 f^N_{2,i}=f^N_{2,i}\ , \qquad\qquad \Theta^m f^N_{m,j}=f^N_{m,j} \, .
\ee
Any vector in $N^*$ can be expressed as $\sum a_i f^N_{2,i}+\sum b_j f^N_{m,j}$ modulo $N$, where $a_i=\{0,1\}$ 
and $b_j=\{0,1,\ldots,m-1\}$. Let $f^I_{2,i}$ and  $f^I_{m,j}$ be the corresponding correlated  $\bz_2$ and  
$\bz_m$ generators of $I^*/I$. The vectors of $\Gamma$ are then of the form
\vs{-10}\be\label{Gamma}
\Gamma=\sum_{i=1}^{\ell_2} a_i( f^I_{2,i}, f^N_{2,i}) +\sum_{j=1}^{\ell_m} b_j (f^I_{m,j}, f^N_{m,j})+ (I,N)\ ,\;\; 
a_i=\{0,1\}\ ,\;\; b_j=\{0,1,\ldots,m-1\}\ .\vs{-10}
\ee
The preceding arguments also indicate that the sublattices $I_m$ and $I_2$ have the form
\be\label{I2Im}
 I_m = (I, N_m) + \sum_{j=1}^{\ell_m} b_j (f^I_{m,j}, f^N_{m,j}) \ ,\qquad\qquad 
 I_2 = (I, N_2) + \sum_{i=1}^{\ell_2} a_i (f^I_{2,i}, f^N_{2,i})\ ,
 \ee
where $N_m$ and $N_2$ consist of vectors in $N$ that are respectively invariant under $\Theta^m$ and $\Theta^2$. The corresponding 
dual lattices $I^*_m$ and $I^*_2$ will be the projection of $\Gamma$ into the respective invariant directions.

The relevant projectors on the invariant subspaces are
\be\label{proj}
\Pi_m= \frac{1}{2}(1+\Theta^m)\ ,\qquad\qquad\Pi_2= \frac{1}{m}(1+\Theta^2+\Theta^4+\ldots+\Theta^{2(m-1)})\, .
\vs{-10}
\ee
It is easy to show that $f^N_{m,j}$ and  $f^N_{2,i}$ have the properties
\be\label{projV2Vm}
\Pi_2f^N_{m,j} = 0\ , \qquad\qquad  \Pi_mf^N_{2,i}=0 \, .
\ee
We now apply the projector $\Pi_2$ on $\Gamma$ given in eq. \eqref{Gamma} and find
\vs{-5}\be\label{I2*}
I_2^* =  \Pi_2\Gamma =\sum_{i=1}^{\ell_2} a_i( f^I_{2,i},f^N_{2,i})+ \sum_{j=1}^{\ell_m} b_j( f^I_{m,j},0)+(I,\Pi_2N)\ ,\vs{-5}
\end{equation}
where we have used \eqref{invV2Vm} and \eqref{projV2Vm}.
Comparing with $I_2$ in eq. \eqref{I2Im} shows that
\vs{-5}\be\label{I2*I2}
I_2^*/I_2 =   \sum_{j=1}^{\ell_m} b_j (f^I_{m,j},0) + (0,\Pi_2N \!\!\!\!\mod N_2)\ .\vs{-5}
\ee
Note that these $\bz_m$ classes are uncorrelated and that the $\bz_m$ classes coming from $N$-directions are just 
$\Pi_2 N$  modulo $\Theta^2$-invariant part of $N$. 
We note that the invariant lattice under $\Theta^{2l}$, $1<l<m$, is also $I_2$.

Let us next apply the projector $\Pi_m$ to $\Gamma$ given in eq. \eqref{Gamma}. 
Using \eqref{invV2Vm} and \eqref{projV2Vm} we obtain
\vs{-5}\be\label{Im*}
I_m^* =  \Pi_m\Gamma = \sum_{i=1}^{\ell_2} a_i( f^I_{2,i},0) + \sum_{j=1}^{\ell_m} b_j (f^I_{m,j}, f^N_{m,j})+(I,\Pi_mN)\ .\vs{-5}
\ee
From the form of $I_m$ given in eq. \eqref{I2Im} we then conclude that
\vs{-5}\be\label{Im*Im}
I_m^*/I_m =   \sum_{i=1}^{\ell_2} a_i (f^I_{2,i},0) + \Big(0,\Pi_mN \!\!\!\!\mod N_m \Big)\ .
\ee
Again these are uncorrelated $\bz_2$ classes. We also observe that the $\bz_2$ classes that come from $N$-directions are 
just $\Pi_m N$ modulo $\Theta^m$-invariant part of $N$. 
Moreover, the length square of the latter $\bz_2$ classes are integers. This is
\be\label{Z2Nm}
\Big(\frac{1}{2}(1+\Theta^m)P\Big)^2=P\cdot\frac{1}{2}(1+\Theta^m)P= \frac12(P^2+P\cdot\Theta^m P) \in \bz\quad {\forall}\;P \in N\ .
\ee
In the last step we have used that $P\cdot\Theta^m P$ is even $\forall \;P \in N$, which in turn can be shown
starting with $P\cdot(1+\Theta+\Theta^2+\ldots+\Theta^{2m-1})P=0$.
We will use the property \eqref{Z2Nm} in section \ref{section_Z} when we study the modular invariance of the asymmetric 
orbifold partition function.

To summarize, the invariant lattices $I$, $I_2$ and $I_m$ have discriminant groups
\be\label{discIs}
I^*/I\cong\bz_2^{\ell_2}\times\bz_m^{\ell_m}\ ,\qquad\qquad I_2^*/I_2\cong\bz_m^{\ell_m+\ell_m^N}\ ,\qquad\qquad
I_m^*/I_m\cong\bz_2^{\ell_2+\ell_2^N}\ .
\ee
Here $\ell_2^N$ is the number of $\bz_2$ generators along $\Theta^m$-invariant directions of $N$.
Similarly, $\ell_m^N$ is the number of $\bz_m$ generators arising from $\Theta^2$-invariant  directions along $N$.

An important piece of the partition function will be the degeneracy factors in the $g^k$-twisted sectors defined as
\be\label{defFs}
\cF_k :=\sqrt{\frac{{\rm det}^{'}(1-\Theta^k)}{|I_k^*/I_k|}} \, .
\ee
These factors have been shown to be integers \cite{Lepowsky, Narain:1986qm}. From the eigenvalues of the automorphism in
\eqref{ta} and the quotient groups in eq. \eqref{discIs} we readily obtain
\be\label{Fs}
\cF_k  =\left\{
\begin{array}{ll}
m^{\frac12(n_2-\ell_m)} \op 2^{\frac12(2n_m-\ell_2)}, & \ k=1 \\[1mm]
m^{\frac12(n_1 + n_2 - \ell_m - \ell^N_m)}, & \ k=2 \\[1mm]
2^{\frac12((m-1)n_1 + 2 n_m - \ell_2 - \ell^N_2)}, & \ k=m
\end{array}
\right.
.
\ee
The fact that the $\cF_k$ are integers then implies
\vs{-5}
\be\label{Fint}
\begin{array}{lll}
n_2-\ell_m\in2\bz_{\ge0}\ , &\quad 2n_m-{\ell_2}\in2\bz_{\ge0}\ , &\quad \ell_2\in2\bz_{\ge0}\ ,\\[3pt]
n_1 + n_2 - \ell_m - \ell^N_m \in2\bz_{\ge0}\ , 
&\quad (m-1)n_1 + 2 n_m - \ell_2 - \ell^N_2 \in2\bz_{\ge0}\ , &\quad \ell^N_2\in2\bz_{\ge0}\, .
\end{array}
\ee
Notice also that if $n_2=0$ then necessarily $\ell_m=0$, and if $n_m=0$ then necessarily $\ell_2=0$.
All these conditions are satisfied by the Leech lattice automorphisms considered in this work.

\section{Partition function of the  \texorpdfstring{$\bt^{D}/\bz_{2m}$}{TEXT} asymmetric orbifolds}\label{section_Z}

We want to study heterotic asymmetric orbifolds $\bt^{D}/\bz_{2m}$ preserving all sixteen supercharges.
As we already explained, in this case the $\bz_{2m}$ automorphism $\Theta$ of $\tDl$ leaves invariant all $D$ right directions 
and acts non-trivially on $s$ left directions. This further means that the orbifold generator, denoted $g$, does not act on the right-moving 
internal bosons, nor on right-moving fermions in order to maintain world-sheet supersymmetry. The action of $g$ on the 
left-moving internal bosons is instead dictated by the eigenvalues of $\Theta$. 
The action on $P \in \Gamma$ depends on $\Theta$ and additional phases as will be discussed shortly. 
As in the previous section we take $m > 2$ to be prime.

To describe the propagation of strings in orbifolds we consider the partition function, or one loop vacuum amplitude, 
that may be written as $Z=\int_{\mathfrak F}\,  \frac{d^2\tau} {\tau_2^2} \, \cz$, where $\mathfrak F$ is the fundamental 
region of $SL(2,\bz)$. In the orbifolds preserving all supersymmetries the integrand factorizes as
\be\label{genpf}
\cz (\tau,\bar \tau) = \cz_X (\tau,\bar \tau)  \ \cz_\psi(\bar\tau)  \ \cz_{\Gamma}(\tau, \bar\tau) \ ,
\ee
where $\cz_X$ and $\cz_\psi$ are respectively the contributions of the uncompactified light-cone world-sheet bosons and the 
right-moving world-sheet fermions given by
\vs{-5}\be
\cz_X(\tau,\bar \tau) =\frac1{\left(\sqrt{\tau_2} \eta \bar\eta\right)^{8-D}}\ , \quad
\cz_\psi(\bar\tau) = \frac1{2\op \bar\eta^4} (\bar\vartheta_3^4 - \bar\vartheta_4^4 - \bar\vartheta_2^4 + 
\bar\vartheta_1^4 ) =\frac{1}{\bar\eta^4} \left(\sum_{r \in V}-\sum_{r \in Sp}\right)\opm \bar q^{\frac12 r^2} \ .
\label{zxzpsidef}
\ee
The conventions for the Jacobi $\vartheta$-functions are those of \cite{Blumenhagen:2013fgp}. 
The $\cz_{\Gamma}(\tau, \bar\tau)$ due to the internal bosons in the $(16+D,D)$ lattice $\Gamma$  can be written as
 \be\label{fullz}
 \cz_\Gamma(\tau,\bar \tau) = \sum_{\ell=0}^{2m-1} \left[ \frac1{2m} \sum_{k=0}^{2m-1} Z(g^\ell, g^k) \right]\ ,
 \ee
where $Z(g^\ell, g^k) = \text{Tr}_{\ch_\ell}\!\! \left(g^k \, q^{L_0} \, \bar q^{\bar L_0}\right)$, and $\ch_\ell$ is the Hilbert 
space of the $g^\ell$-twisted sector. The sum over $\ell$ is a sum over untwisted ($\ell=0$) and twisted ($\ell>0$) sectors 
whereas the sum over $k$ enforces the orbifold projection.
The twisted sector contributions $Z(g^\ell, g^k)$, $\ell>0$, may be computed from untwisted sector terms $Z(\uno,g^k)$
by applying modular transformations generated by the $S$-transformation $\tau \to -1/\tau$ and the $T$-transformation
$\tau \to \tau + 1$. Validity of operator interpretation might impose extra constraints on the conformal field theory.

In the following subsections we compute the explicit form of $Z(g^\ell,g^k)$ in untwisted and twisted sectors. 
From now on we will focus on automorphisms which act on {\it all} left-moving directions, i.e. $s=16+D$, and leave invariant a purely 
right lattice with signature $(0,D)$. This is a necessary condition for islands, as they do not have any vector multiplets in the spectrum. 
In this section we will provide general expressions for the partition function for this choice of $s$.
Nonetheless, in appendix \ref{apprr}, we construct two asymmetric orbifolds with rank reduction, for $\bt^4/\bz_2$ and $\bt^4/\bz_6$ 
heterotic compactifications, in which the invariant lattice has both left and right directions. 
Since in all examples $D$ is an even number, hereafter we will set $D=2d$ to simplify expressions.

\subsection{Untwisted sector}\label{subsection_unt}

The purely untwisted term is
\be\label{z11}
Z(\uno,\uno)= \frac1{\eta^{16+2d} \bar\eta^{2d}}\, \sum_{P\, \in \, \Gamma} q^{\frac12 P_L^2} \bar q^{\frac12 P_R^2} \ ,\vs{-5}
\ee
where $\Gamma$ is the $(16+2d,2d)$ even self-dual compactification lattice.

Let us know consider $Z(\uno,g)$. We will assume that the action of $g$ on $P \in \Gamma$ is such the trace over the untwisted 
Hilbert space is as in eq. \eqref{sch1g}, where $v$ is a constant shift vector along the $I$ directions. We argued before that this
might not lead to a modular invariant partition function because it is not guaranteed that $2 m P^2$ is even for all $P$ in $I^*$.
The problem arises when there are $\bz_2$ classes of $I^*/I$ with lengths squared $1/2$ mod integers. 
In this case we have shown that there exists a constant vector $w$, with $w \in I^*$ and $2 w \in I$, such that 
$2 m P^2 = 2 P \cdot w \, \text{mod}\, 2$. Now, since the phase that appears upon $\tau \to \tau + 2m$ inside the sum in
\eqref{schg1} is actually $e^{2 i\pi m (P+v)^2}$, we observe that the momentum dependent piece is nothing but
$e^{2 i\pi P\cdot (2mv + w)}$. To ensure that this piece is equal to one for all $P \in I^*$ it suffices to require
\be\label{condvw}
(2mv + w) \in I \, .
\ee
In the following we will impose this condition on the shift $v$. Notice that $v$ has order $4m$, i.e. $4m v \in I$.

The action of $g$ on the left-moving world-sheet bosons along the $N$ directions is
\be
g X_a = e^{2\pi i  t_a} X_a \ ,\qquad\qquad a=1, \ldots, 8+d  \quad (\text{no sum in}\  a) \ ,
\ee
where $X_a$ are complex, and $t_a$ are given in eq. \eqref{ta}. 
It is well known that the contribution of the $X_a$ to the partition function can be expressed in terms of 
Jacobi Theta functions, see e.g. \cite{Blumenhagen:2013fgp}. 
As discussed before, taking into account the action of the orbifold on the lattice then leads to
\vs{-5}\be
Z(\uno,g)= \frac1{\eta^{16+2d} \bar\eta^{2d}}\, 
\prod_{a=1}^{8+d} \frac{2 \sin (\pi  t_a)\,  \eta^3}
{\vtgen{\frac12}{\frac12 - t_a} }
\sum_{P\, \in \, I} q^{\frac12 P_L^2} \bar q^{\frac12 P_R^2} \, e^{2\pi i P \cdot v}\ .
\label{z1g}
\ee
To avoid confusion with the notation we remark that scalar products are with Lorentzian metric, unless the subscripts 
$L$ or $R$ are explicitly written. For instance, $P^2 = P_L^2 - P_R^2$, whereas \mbox{$P\cdot v = P_L \cdot v_L - P_R\cdot v_R$}.

In $Z(\uno,g^2)$ the momentum sum will be over $I_2$, which is the lattice invariant under $\Theta^2$.  
The action of $g^2$ on $\Gamma$ also involves a phase that is fixed by consistency with modular invariance.
Specifically, in subsection \ref{subsection_g2} we will apply a $ST^2S$ transformation to the above $Z(\uno,g)$ to
obtain $Z(g^2,g^{-1})$. The result in eq. \eqref{Zg2gm1} shows that the lattice momenta are shifted by $2v+w$.
Hence, according to the operator interpretation, the same shift must also occur in $Z(g^2,\uno)$, which  
follows by applying an $S$ transformation to $Z(\uno,g^2)$. The conclusion is that $g^2$ must pick up a phase
$e^{2 i \pi P\cdot(2v +w)}$ when acting on $P \in I_2$.
Altogether, for the $Z(\uno,g^2)$ term the result is 
\be\label{z1g2}
Z(\uno,g^2)= \frac1{\eta^{16+2d} \bar\eta^{2d}}\, \prod_{\substack{a=1 \\ 2t_a \neq 1}}^{8+d}
\frac{2 \sin (2\pi  t_a)\,  \eta^3}{\vtgen{\frac12}{\frac12 - 2t_a} }
\sum_{P\, \in \, I_2} q^{\frac12 P_L^2} \bar q^{\frac12 P_R^2} \, e^{2\pi i P \cdot (2v+w)} \ .
\ee
Notice that now the product excludes the $a$ for which $2t_a=1$. Along such a direction the complex boson just contributes $1/\eta^2$. 

More generally, in $Z(\uno,g^\ell)$ there will be a lattice sum over $P \in I_\ell$, which by definition is the invariant lattice
under $\Theta^\ell$. From modular transformations we infer that the $g^\ell$ action on $P \in I_\ell$ includes the phase
$e^{2\pi i P \cdot(\ell v+w\delta_{\ell,2\!\!\!\mod\!\!4})}$. 
We may then write the general expression
\be\label{z1gl}
Z(\uno,g^\ell)= \frac1{\eta^{16+2d} \bar\eta^{2d}}\, \prod_{\substack{a=1 \\ \ell t_a \notin\bz}}^{8+d}
\frac{2 \sin (\ell\pi  t_a)\,  \eta^3}{\vtgen{\frac12}{\frac12 - \ell t_a} }
\sum_{P\, \in \, I_\ell} q^{\frac12 P_L^2} \bar q^{\frac12 P_R^2} \, e^{2\pi i P \cdot (\ell v+w\delta_{\ell,2\!\!\!\!\!\mod\!4})} \ .
\ee
The structure of the invariant lattices $I_\ell$ is explained in subsection \ref{subsection_I}. 
Recall that $I_\ell=I_1$ for $(\ell,2m)=1$, whereas $I_{2j}=I_2$ for $1 < j < m$.

For $Z(\uno,g^m)$ the oscillator contribution is actually simpler because the non-trivial eigenvalues of $\Theta^m$ are $-1$. We find
\be\label{z1gm}
Z(\uno,g^m)= \frac1{\eta^{16+2d} \bar\eta^{4}}\,\left(\frac{2 \eta^3}{\vartheta_2}\right)^{\!\!\! \frac{\msm{s}_m}{2} }\!\!
\sum_{P\, \in \, I_m} q^{\frac12 P_L^2} \bar q^{\frac12 P_R^2} \, e^{2\pi i P \cdot mv} \ .
\ee
Here $\sbig_m$ is defined to be the number of $-1$ eigenvalues of $\Theta^m$. It is explicitly given by
\be\label{bigs}
\sbig_m = (m-1) n_1+ 2 n_m\ , 
\ee
as follows from the general form of the eigenvalues in \eqref{ta}.

Some additional comments are in order.
Upon an $S$-transformation, we learn that in the $g^{2(2j-1)}$-twisted 
sector, the lattice that appears is $I_2^*$ shifted by $2(2j-1)v + w$, while in the $g^{2(2j)}$-sector the shift will be just $2(2j)v$, 
for some positive integer $j$. Consider now the OPE of a state in the $g^{2l}$-twisted sector with another state in
the inverse $g^{2(m-l)}$-twisted sector. If $l$ is even, then $(m-l)$ is odd, and vice versa. Hence, one of these two twisted states will 
come with $w$, and the total shift will be $2m v+w$, which is in $I$, as it should since the OPE gives a state in the untwisted sector.

\subsection{\texorpdfstring{$g$-}{TEXT}twisted sector and modular invariance}\label{subsection_g}

Applying an $S$-modular transformation, followed by $T$-transformations, to $Z(\uno,g)$ we obtain the twisted sector terms
\be
Z(g,g^j)= \frac{\cF_1}{\eta^{16+2d} \bar\eta^{2d}}\, 
\prod_{a=1}^{8+d} \frac{ e^{i\pi(\frac12 - t_a)} e^{i\pi j t_a^2} \eta^3}
{\vtgen{\, \frac12-t_a}{\frac12 -j t_a} }
\sum_{P\, \in \, I^*}  \bar q^{\frac12 (P+v)_R^2} \op e^{i\pi j (P+v)^2} \ , 
\label{zggja}
\ee
where we have taken into account that $I$ has only right-moving directions. The degeneracy factor $\cF_1$ is defined
in \eqref{defFs}, see also \eqref{Fs}.
To simplify the expressions, it is convenient to define
\vs{-5}\be\label{thatdef}
\hvtgen{\,\alpha}{\beta} = e^{-2 i\pi  \alpha \beta} \vtgen{\,\alpha}{\beta} \ .\vs{-5}
\ee
The advantage of using the $\widehat\vartheta$'s is that  the overall phase depending on the eigenvalues 
of $\Theta$ simplifies considerably. Indeed, eq. \eqref{zggja} can be rewritten as
\be\label{zggjb}
Z(g,g^j)= \frac{\cF_1 \op e^{2\pi ij E_1} }{\eta^{16+2d} \bar\eta^{2d}}\, 
\prod_{a=1}^{8+d} \frac{\eta^3}{\hvtgen{\, \frac12-t_a}{\frac12 -j t_a} }
\sum_{P\, \in \, I^*} \bar q^{\frac12 (P+v)_R^2} \op e^{i\pi j (P+v)^2} \ . 
\ee
Here $E_1$ is the zero point shift given by
\be\label{ezero1def}
E_1 = \frac12\sum_{a=1}^{8+d} t_a(1-t_a) \, .
\ee
From the explicit form of the $t_a$'s in\eqref{ta} we easily compute
\be\label{ezero1}
E_1=\frac{(m-1)}{48m}\Big((2m-1)n_1+2(m+1)n_2\Big)+\frac{n_m}8 \, ,
\ee
which will be needed later. Notice that $(n_1 + n_2)(m-1) + 2 n_m=16+2d$.

Modular invariance requires that $Z(g,g^{2m})=Z(g,\uno)$. 
Now, the result in \eqref{p2wcond} informs us that
\be
e^{ i \pi 2 m (P+v)^2}  =  e^{2 i \pi P \cdot(2mv+w)} e^{2 i \pi m v^2}, \quad \forall \, P \in I^* \ .\vs{-5}
\ee
Thus, as already argued, the momentum dependent piece drops out imposing condition \eqref{condvw}, namely
$(2mv + w) \in I$.
In the end, requiring $Z(g,g^{2m})=Z(g,\uno)$ yields the level matching condition
\be
\label{LMCg}
4m E_1 + 2m v^2 \in 2\bz \ .
\ee
Note in particular that when $4mE_1$ is odd, the trivial shift $v=0$ is not a solution.

\subsection{\texorpdfstring{$g^{2j+1}$-}{TEXT}twisted sectors \texorpdfstring{$2j+1\ne m$}{TEXT}}\label{subsection_godd}

Let us now look at the $g^{(2j+1)}$-twisted sectors, with $j\!\!=\!\!\{0, \ldots, m-1\}$, excluding $j\!=\!\frac{m-1}2$ 
(the case $j\!=\!0$ was considered in the previous subsection). To simplify the expressions, in the following we denote 
$\ell=(2j+1)$. Similar to the $g$-twisted sector, these sectors have a single orbit under the $T$ transformation $\tau \to \tau+1$.

The starting point is $Z(\uno,g^\ell)$ obtained from eq. \eqref{z1gl}. 
It is helpful to define
\be\label{exfac2}
\prod_{a=1}^{8+d}  2\sin (\pi \ell t_a) := e^{i\pi \mu} \, \bigg|\prod_{a=1}^{8+d}  2\sin (\pi\ell t_a) \bigg| =
e^{i\pi \mu}\, \sqrt{{\rm det}^{'}(1-\Theta^\ell)} = e^{i\pi \mu}\, \sqrt{{\rm det}^{'}(1-\Theta)}
\, .
\ee
where the phase $e^{i\pi \mu}$ can be easily evaluated and clearly equals $\pm 1$.
Applying an $S$-transformation $\tau \to -1/\tau$ to $Z(\uno,g^\ell)$ leads to
\be
Z(g^\ell,1)= \frac{\cF_1 \op e^{i\pi \mu}}{\eta^{16+2d} \bar\eta^{2d}}\, 
\prod_{a=1}^{8+d} \frac{ e^{i\pi(\frac12 -\ell  t_a)}\, \eta^3}
{\vtgen{\, \frac12-\ell t_a}{\frac12} }
\sum_{P\, \in \, I^*} \bar q^{\frac12 (P+\ell v)_R^2} \ .\vs{-5} 
\label{zgl1a}
\ee
Notice that the degeneracy factor $\cF_1$ in the $g^\ell$-sector is the same as in the $g$-sector.

To extract the actual overall factor it is convenient to define
\be\label{tlta}
\widetilde{\ell t_a} := \ell t_a - \lfloor \ell t_a \rfloor \ .
\ee
Recall that $\lfloor x \rfloor$ is the greatest integer less than or equal to $x$, so that $0<\widetilde{\ell t_a} < 1$.
Using the definition \eqref{thatdef}, we arrive at the final expression
\be\label{zgl1c}
Z(g^\ell,\uno)= \frac{\cF_1 }{\eta^{16+2d} \bar\eta^{2d}}\, 
\prod_{a=1}^{8+d} \frac{ \eta^3}
{\hvtgen{\, \frac12-\widetilde{\ell t_a}}{\frac12} }
\sum_{P\, \in \, I^*} \bar q^{\frac12 (P+\ell v)_R^2} \ .\vs{-10} 
\ee
The modified $\vartheta$ functions directly give the oscillator expansion in positive powers of $q$, assuring
that the overall factor is the degeneracy $\cF_1$.

Other terms in the $g^\ell$-sector are obtained performing $T$-transformations. For example
\be\label{zglgl}
Z(g^\ell,g^\ell)= \frac{\cF_1 \op e^{2 \pi iE_\ell}}{\eta^{16+2d} \bar\eta^{2d}}\, 
\prod_{a=1}^{8+d} \frac{ \eta^3}{\hvtgen{ \frac12-\widetilde{\ell t_a}}{\frac12 -\widetilde{\ell t_a}} }
\sum_{P\, \in \, I^*} \bar q^{\frac12 (P+\ell v)_R^2}  \op e^{i\pi (P+\ell v)^2}  \ .\vs{-5} 
\ee
Here $E_\ell$ is the zero point energy in the $g^\ell$-sector, namely 
$E_\ell=\frac12 \sum\limits_{a=1}^{8+d} \widetilde{\ell t_a} (1 - \widetilde{\ell t_a})$. 
One can show that $E_\ell=E_1$. 

Performing a $\tau \to \tau + 2m$ transformation, we find the modular invariance condition
\be
4m E_1 + 2m \ell^2 v^2 \in 2\bz \ .
\label{modinvgl}
\ee
Given the general form of the eigenvalues $t_a$ \eqref{ta}, and the fact that $\ell=(2j+1)$, it follows that 
the above modular invariance condition reduces to that in  the $g$-sector, i.e. $4m E_1 + 2mv^2 \in 2\bz$.

\subsection{\texorpdfstring{$g^{2j}$}{TEXT}-twisted sectors}\label{subsection_g2}

We will compute the $g^2$-sector partition function explicitly. Other even-twisted sector partition functions are computed in a similar 
way, and the corresponding expressions will be presented at the end. It is convenient to consider separately the orbits under 
$T$-transformations of $Z(g^2,\uno)$ and $Z(g^2,g^{-1})$. 

\subsubsection{Orbit \texorpdfstring{$Z(g^2,\uno)$}{TEXT}}\label{subsub_g21}

Applying modular transformations to $Z(\uno,g^2)$ in \eqref{z1g2} gives
\be
Z(g^2,g^{2k})= \frac{\cF_2\op e^{i\pi 2k E_2} }{\eta^{16+2d} \bar\eta^{2d}}\, 
\prod_{\substack{a=1 \\ 2t_a \neq 1}}^{8+d}\frac{\eta^3}
{\hvtgen{\!\!\!\frac12-2t_a}{\frac12 -2k t_a} }
\sum_{P\, \in \, I_2^*} q^{\frac12 (P+2v+w)_L^2} \op \bar q^{\frac12 (P+2v+w)_R^2} \op e^{i\pi k (P+2v+w)^2} \ ,
\label{zg2geven}
\ee
where the degeneracy factor $\cF_2$ is defined in eq. \eqref{defFs}, see also \eqref{Fs}.
The zero point shift $E_2$ is defined by
\be\label{ezero2}
E_2 = \frac12\sum_{a=1}^{8+d} 2t_a(1-2t_a) = \frac{(m^2-1)}{24m}(n_1+n_2)\, ,
\ee
where for the second equality we inserted the specific $t_a$'s in\eqref{ta}. 

Modular invariance requires $Z(g^2,g^{2m})=Z(g^2,\uno)$. Consider eq. \eqref{zg2geven} with $k=m$. We first show
\be
e^{i \pi m (P+2v+w)^2} = 
e^{i \pi m (2v+w)^2}\quad\forall\; P\in I_2^*\ .
\ee
To prove this, note that $I_2^*$ has uncorrelated $\bz_m$ classes -- see eq. \eqref{I2*I2}. 
Thus, $mP^2\in2\bz$ for all $P\in I_2^*$. Furthermore, $P\cdot\,2mv\in\bz$ and $P\cdot w\in\bz$ because $2mv$ and 
$w$ are in $\bz_2$ classes of $I^*$. 
Therefore, modular invariance results in the level matching condition
\be\label{LMCg2}
2m E_2 + m (2v+w)^2 \in 2\bz \ .
\ee
Using conditions \eqref{condvw} and \eqref{LMCg}, together with eq. \eqref{ezero2}, one can show that the above equation reduces to
\be\label{LMCg2_ii}
w^2+ \frac{\epsilon-1}2n_1+n_m \in 2\bz \ ,\vs{-5}
\ee
where $m=\epsilon\!\mod4$ and $\epsilon^2=1$. Note that the above requires $w^2\in\bz$. In appendix \ref{appC2} we show that
actually $w^2 \in \bz$ is a property of a $(0,2d)$ lattice invariant under a $\bz_{2m}$ automorphism of $\Gamma$.  

\subsubsection{Orbit \texorpdfstring{$Z(g^2,g^{-1})$}{TEXT}}\label{subsub_g2g1}

We apply an $S$-transformation to $Z(g,g^2)$ in eq. \eqref{zggja}. The lattice part of the $Z(g^2,g^{-1})$ orbit is 
more involved and is computed in appendix \ref{app_g2}, via performing Poisson resummation -- the final result is 
eq. \eqref{g2alattb}. Combining with the oscillator piece, we obtain
\be
\begin{split}\label{Zg2gm1}
Z(g^2,g^{-1}) &= e^{i\frac{\pi}{2}(8+d)}e^{i \pi \theta_m^1}\op \frac{\cF_2'\op e^{-i\pi E_2} }{\eta^{16+2d} \bar\eta^{2d}}\, 
\left(\frac{2 \eta^3}{\vartheta_2}\right)^{\!\! n_m }\prod_{\substack{a=1 \\ 2t_a \neq 1}}^{8+d}
\frac{\eta^3}{\hvtgen{\, \frac12-2t_a}{\frac12 + t_a} } \\[-1mm]
&\times \op e^{2\pi iv^2} \op \sum_{b_i=0}^{m-1} \sum_{P\, \in \, I} \bar q^{\frac12 (P+2 b_i f_i + 2v+w)_R^2} \op 
e^{-2i\pi (P+2 b_i f_i + 2v+w)\cdot v} \op e^{-2\pi i(b_i f_i)^2}\ , 
\end{split}
\ee
where the degeneracy factor is
\be\label{F2p}
\cF_{2}'=m^{\frac12(n_2-\ell_m)}\ .\vs{-5}
\ee
To simplify expressions we have defined
\vs{-10}\be\label{bifi}
b_i f_i := \sum_{i=1}^{\ell_m} b_i f_i \ ,\qquad\qquad b_i=\{0,\ldots,m-1\}\ ,\vs{-5}
\ee
where $f_i$ are the $\ell_m$ generators of $\bz_m$ classes of $I^*/I$. The phase $e^{i \pi \theta_m^1}$ 
is defined through
\vs{-10}\be
C_m(1) = e^{i \pi \theta_m^1}\,\sqrt{m^{\ell_m}} = \sum_{b_i=0}^{m-1} e^{2i\pi(b_i f_i)^2}\vs{-5}
\ee
(see eq. \eqref{Cma} and appendix \ref{appCma} for details). 

Let us emphasize that to obtain the lattice part of $Z(g^2,g^{-1})$ we performed a Poisson resummation over a lattice
larger than $I$, as explained in appendix \ref{app_g2}. There is an alternative way of obtaining the lattice part that is briefly 
discussed in appendix \ref{app_Lsum}. However, the procedure followed here has the great advantage of making the spectrum 
manifest. Concretely, in eq. \eqref{Zg2gm1} we clearly see that the momenta that enter are  $(P'+2 v+ w)$ with 
$P' \in \bz_m$  classes of $I^*$. Moreover, \eqref{I2*} shows that such $P'$ are precisely the $g$-invariant part of 
$I_2^*$, which is the momentum lattice that appears in $Z(g^2,\uno)$. This is a basic requirement for the operator interpretation
of $Z(g^2,g^{-1})$ as arising from $\text{Tr}_{\ch_2}\!\! \left(g^{-1} \, q^{L_0} \, \bar q^{\bar L_0}\right)$, since
the trace only keeps the $g$-invariant states in the $\ch_2$ Hilbert space that appears in $Z(g^2,\uno)$.

After a $\tau \to \tau + k$ transformation we find 
\be\begin{split}
Z(g^2,g^{2k-1}) &=  e^{i\frac{\pi}{2}(8+d)}\op e^{i \pi \theta_m^1}\op\frac{\cF_2'\op e^{i\pi (2k-1) E_2} }{\eta^{16+2d} \bar\eta^{2d}}\, 
\left(\frac{2 \eta^3}{\vartheta_2}\right)^{\!\!n_m }\prod_{\substack{a=1 \\ 2t_a \neq 1}}^{8+d}
\frac{\eta^3}{\hvtgen{\, \frac12-2t_a}{\frac12 -(2k-1) t_a} } \\[1mm]
&\hspace*{-1.5cm} \times \op e^{2\pi iv^2} \op 
\sum_{b_i=0}^{m-1} \sum_{P\, \in \, I} \bar q^{\frac12 (P+2 b_i f_i + 2v+w)_R^2} \op e^{i\pi k (P+2 b_i f_i + 2v+w)^2}
\op e^{-2i\pi (P+2 b_i f_i + 2v+w)\cdot v}  \op e^{-2\pi i(b_i f_i)^2} \, .
\end{split}\label{zg2godd}
\ee
Recall that $I$ has only right-moving directions and that scalar products and norms are computed with Lorentzian metric.

Applying $\tau \to \tau + 2m$ to $Z(g^2, g^{-1})$ does not give rise to additional modular invariance conditions beyond
eq. \eqref{LMCg2_ii} obtained from $Z(g^2,\uno)=Z(g^2,g^{2m})$. The basic reason is that, as mentioned above, the momenta  
that enter in \eqref{Zg2gm1} are a subset of $I_2^*$, which is the momentum lattice in $Z(g^2,\uno)$.

It is important to stress that e.g. $Z(g^2,g^{-1})$ is obtained by a chain of modular transformations starting from $Z(\uno,g)$.
Thus, to arrive at \eqref{Zg2gm1} we only needed to know that the rotation part of $g$ restricts momenta to lie in the invariant
lattice $I$ and that $g|P\rangle = e^{2 i \pi P\cdot v} |P\rangle $ for all $P \in I$. On the other hand, $Z(g^2,g^{-1})$ must properly
include the action of $g$ in the Hilbert space of the $g^2$-twisted sector which we can read from $Z(g^2,\uno)$ in eq. \eqref{zg2geven}.
As observed already, the  momenta entering in $Z(g^2, g^{-1})$ are $(P'+2 v+ w)$ with $P'$ belonging to the
$g$-invariant subset of the momentum lattice $I_2^*$ in $Z(g^2,\uno)$. Now we want to remark 
that the fact that the momenta in $Z(g^2,g^{-1})$ are shifted
by $(2v+w)$ tells us that the same shift should occur in $Z(g^2,\uno)$, which in turn is obtained by $S$-transformation
from $Z(\uno,g^2)$. This then requires that $g^2|P\rangle = e^{2 i \pi P\cdot(2v +w)} |P\rangle $
for all $P$ in $I_2$, which is the lattice invariant under $\Theta^2$.  
We thus see that we are forced to include the additional shift by $w$ in the action of $g^2$ on $I_2$.

\subsubsection{Other even-twisted sectors}

A similar analysis may be done for other $g^{2j}$-twisted sectors, $1<j<m$. 
For the $Z(g^{2j},\uno)$ orbit we find
\be\label{zgevengeven}
 Z(g^{2j},g^{2jk})= \frac{\cF_2\op e^{i\pi 2k E_2} }{\eta^{16+2d} \bar\eta^{2d}}\, 
\prod_{\substack{a=1 \\ 2t_a \neq 1}}^{8+d}\frac{\eta^3}
{\hvtgen{\!\!\!\frac12-\widetilde{2jt_a}}{\frac12 -k \widetilde{2jt_a}} }
\sum_{P\, \in \, I_2^*} q^{\frac12 (P+2jv+jw)_L^2} \op \bar q^{\frac12 (P+2jv+jw)_R^2} \op e^{i\pi k (P+2jv+jw)^2} \ .
\ee
For the $Z(g^{2j},g^{-1})$ orbit the result is
\be\begin{split}
\!\!Z(g^{2j},g^{2jk-1}) &\!=  e^{i\frac{\pi}{2}(8+d+(j-1)n_m)}\op e^{i \pi \theta_m^j}\op\frac{\cF_2'\op 
e^{i\pi 2k E_2} }{\eta^{16+2d} \bar\eta^{2d}}\!
\prod_{\substack{a=1 \\2t_a\ne1}}^{8+d} e^{i\pi\lfloor2j t_a\rfloor}
e^{i\pi(\widetilde{2j t_a}-\lfloor2j t_a\rfloor-1)t_a}\\[1mm]
&\qquad\qquad\qquad\qquad\;\;\times\bigg(\frac{2\eta^3}{\vartheta_2} \bigg)^{n_m}\prod_{\substack{a=1 \\2t_a\ne1}}^{8+d}
\frac{\eta^3}{\hvtgen{\frac12- \widetilde{2j t_a}}{\frac12-k\,\widetilde{2j t_a}+ t_a} }\\[1mm]
&\hspace*{-2.65cm} \times e^{2\pi ijv^2} \!
\sum_{b_i=0}^{m-1} \sum_{P\, \in \, I} \bar q^{\frac12 (P+2j b_i f_i + 2jv+jw)_R^2} \op e^{i\pi k (P+2 j b_i f_i + 2j v+j w)^2}
\op e^{-2i\pi (P+2j b_i f_i + 2jv+jw)\cdot v}  \op e^{-2\pi ij(b_i f_i)^2} \, .
\end{split}\label{zgevengodd}
\ee
Note that $E_{2j}=E_2$, and that the degeneracy factors $\cF_2$ and $\cF_2'$ are the same as those in the $g^2$ sector. 
The level matching condition is also the same as that of the $g^2$-twisted sector \eqref{LMCg2}.

\subsection{\texorpdfstring{$g^m$}{TEXT}-twisted sector}\label{subsection_gm}

The $g^m$-twisted sector has $m$ $T$-orbits. It is convenient to choose basis $Z(g^m,\uno)$ and 
$(g^m,g^{-(2j+1)})$, with $j=\{0, \ldots,m-1\}$, excluding $2j+1=m$. 
The reason for this choice is that the basis elements are obtained by a single $S$-transformation, either from
$Z(\uno, g^m)$ or from $Z(g^{2j+1}, g^m)$. For each element of the basis we can then
apply a $T$-transformation to derive the remaining $Z(g^m, g^k)$ terms.

\subsubsection{Orbit \texorpdfstring{$Z(g^m,\uno)$}{TEXT}}\label{subsubgm1}

From an $S$-transformation of $Z(\uno,g^m)$ in eq. \eqref{z1gm} we obtain
\be\label{zgm1}
Z(g^m,\uno)= \frac{\cF_m}{\eta^{16+2d} \bar\eta^{2d}}\, 
\left(\frac{\eta^3}{\vartheta_4}\right)^{\!\!\! \frac{\msm{s}_m}{2} }
\sum_{P\, \in \, I_m^*} q^{\frac12 (P+mv)_L^2} \op \bar q^{\frac12 (P+mv)_R^2}\, .
\ee
Here $\sbig_m$ is the number of eigenvalues of $\Theta^m$ equal to $-1$, given in eq. \eqref{bigs}. The degeneracy 
factor $\cF_m$ is defined and evaluated in eqs. \eqref{defFs} and \eqref{Fs}. 

Doing $\tau \to \tau + 1$ gives
\vs{-5}\be
Z(g^m,g^m)= \frac{\cF_m\op e^{2\pi iE_m}}{\eta^{16+2d} \bar\eta^{2d}}\, 
\left(\frac{\eta^3}{\vartheta_3}\right)^{\!\!\! \frac{\msm{s}_m}{2} }\!\!
\sum_{P\, \in \, I_m^*} q^{\frac12 (P+mv)_L^2} \op \bar q^{\frac12 (P+mv)_R^2} \, 
e^{i\pi(P+mv)^2} \, ,
\label{zgmgma}
\ee
where $E_m=\frac{\sbig_m}{16}$ is the zero point shift in the $g^m$-sector. For future use we record the 
explicit value
\be\label{ezerom}
E_m=
\frac1{16}\;\Big((m-1)n_1+2n_m\Big) \, .
\ee
Recall that $n_1$ and $n_m$ refer to multiplicities of $\Theta$ eigenvalues in \eqref{ta}.

Modular invariance requires that $Z(g^m,\uno)$ returns to itself under $\tau \to \tau +2$.
Applying this transformation to  eq. \eqref{zgm1} will clearly produce a phase $e^{2i\pi(P+mv)^2}$ inside the 
sum over $P \in I_m^*$, but the momentum dependent piece equals 1, i.e.
\be\label{auxgm}
e^{2\pi i (P^2+P\cdot 2mv)} = e^{2\pi i P \cdot (w+2mv)}= 1\quad\forall\; P\in I_m^*\ .
\ee
To show this, we recall that $I_m^*$ has uncorrelated $\bz_2$ classes. More precisely, eq. \eqref{Im*Im} tells us
that $P \in I_m^*$ can be decomposed as $P=X+Y$, where $X$ is in a $\bz_2$ class along $I$-directions and $Y$ is in a $\bz_2$ 
class along $g^m$-invariant directions of $N$. Furthermore, in eq. \eqref{Z2Nm} we proved that $Y^2 \in \bz$.
Using also eq. \eqref{w}, and the fact $X\cdot Y=0$, in particular $w\cdot Y=0$, then implies that $P^2=P\cdot w$ mod
integers. The second equality in \eqref{auxgm} follows from $2mv+w\in I$.
In conclusion, $Z(g^m,g^{2m})=Z(g^m,\uno)$ gives the level matching condition
\vs{-5}\be\label{LMCgm}
2 E_m + m^2 v^2 \in \bz \ .\vs{-5}
\ee
It can be shown that this constraint is contained in the level matching condition in the $g$-twisted sector \eqref{LMCg}.

\subsubsection{Orbits $Z(g^m, g^{-(2j+1)})$}\label{subsubgmodd}

To work out the $Z(g^m, g^{-(2j+1)})$ orbits, with $j=0, \ldots, (m-1)$, excluding $j=\frac{m-1}2$, we start with $Z(g^{(2j+1)},\uno)$, 
in which it is important to take into account eq. \eqref{exfac2}. Next we do a \mbox{$T^m$-transformation}, to reach 
$Z(g^{(2j+1)}, g^m)$, after using $g^{2mj}=1$. An \mbox{$S$-transformation} then gives $Z(g^m, g^{-(2j+1)})$. 

Proceeding as explained above leads to
\be\label{zgmgodd}
Z(g^m,g^{-(2j+1)})=  \frac{\cF_1 \op e^{i\pi \mu} }{\eta^{16+2d}}\, 
\prod_{a=1}^{8+d} \frac{ e^{i\pi(1 -m(2j+1) t_a -m(2j+1)^2 t_a^2)}\op \eta^3}
{\vtgen{\, \frac12-mt_a}{\frac12 +(2j+1) t_a} } \;\; Z_{\text{lat}}(g^m, g^{-(2j+1)})\ ,
\ee
where $e^{i\pi \mu}=\pm1$ is the phase defined in \eqref{exfac2}, and $\cF_1$ is given in eq. \eqref{Fs}. 
The lattice part $Z_{\text{lat}}$ is computed in appendix \ref{app_gm} -- eq. \eqref{kmf} is the final result. It is given by
\be\label{zgmgoddlata}
Z_{\text{lat}}(g^m, g^{-(2j+1)}) = \frac{C_2}{\text{vol}_m} \,\frac{e^{i\pi m(2j+1)^2v^2}}{\bar\eta^{2d}} \!\!\!
\hspace*{-2mm} \sum_{\substack{P\, \in \, \bz_2\\ \text{classes of}\, I^*}} \hspace*{-3mm} 
\bar q^{\frac12 (P+mv)_R^2} \op e^{-2i\pi (P+mv)\cdot (2j+1) v} \op e^{-i  \pi\epsilon (P-jw)^2} \ ,
\ee
where ${\rm vol}_m$ and $C_2$ are defined in eqs. \eqref{volm} and \eqref{C2}, and the quantity $\epsilon$ is $\pm1$ according to 
$m=\epsilon \ \text{mod} \ 4$. As usual, norms and scalar products are evaluated with Lorentzian metric. For future purposes 
we record an alternative form of the lattice part
\be\label{zgmgoddlatb}
\hspace*{-3.6mm} 
Z_{\text{lat}}(g^m, g^{-(2j+1)}) \!=\! \frac{C_2}{\text{vol}_m}\op \frac{e^{im \pi (2j+1)^2v^2}}{\bar\eta^{2d}} 
e^{-i\pi \epsilon j^2 w^2} 
\hspace*{-5mm} \sum_{\substack{P\, \in \, \bz_2\\ \text{classes of}\, I^*}} \hspace*{-3mm} \!\!\!
\bar q^{\frac12 (P+mv)_R^2} \op e^{-2i\pi (P+mv)\cdot (2j+1) v} \op e^{-i \epsilon \pi (2j+1)P^2}  ,
\ee
which is obtained using that for all $P \in \bz_2$ classes of $I^*$ there are identities
$e^{2i\pi P^2} = e^{2i \pi P\cdot w}$ and  $e^{4i\pi P^2} =1$.

Before inserting eq.~\eqref{zgmgoddlatb} into \eqref{zgmgodd} to compute the full partition function, we define the phase
\vs{-5}\be\label{phaseDelta}
e^{i\pi \Delta(m,j)} := \prod_{a=1}^{8+d} e^{i\pi(1 -m(2j+1) t_a -m(2j+1)^2 t_a^2)}\ ,
\ee
and the phase $e^{i\pi \nu(m,j)}$ through
\be\label{phasenu}
\prod_{\substack{a=1 \\ mt_a \in \bz}}^{8+d}  -2\sin \big((2j+1)\pi t_a\big) =: e^{i\pi \nu(m,j)} \sqrt{m^{n_2}}\, ,
\ee
where $\nu(m,j)$ equals 0 or 1. Eq. \eqref{phasenu} may be derived using the structure of the eigenvalues given in 
eq. \eqref{ta}. Furthermore, using the eigenvalue structure, it can be shown that 
$e^{i\pi \Delta(m,j)}e^{i\pi \nu(m,j)}e^{i\pi \mu}$ is independent of $j$. We can then evaluate at $j=0$ to obtain
\be\label{delnumu}
e^{i\pi\Delta(m,0)}e^{i\pi \nu(m,0)}e^{i\pi \mu}=e^{-i\frac{\pi}{24}(m-1)(4m-17)n_1}
e^{-i\frac{\pi}{6}(m^2-1)n_2}e^{-i \frac\pi4(3m-4)n_m}\ .\vs{-5}
\ee

We have now laid out all the ingredients that enter in $Z(g^m,g^{-(2j+1)})$. Starting with eq. \eqref{zgmgodd}, we obtain 
\be
\begin{split}
Z(g^m,g^{-(2j+1)})&= \frac{\cF_m'\op e^{i\pi \chi_m}}{\eta^{16+2d}\op  \bar\eta^{2d}}\, 
\prod_{\substack{a=1 \\ mt_a \in \bz}}^{8+d} \frac{ -2\sin \big((2j+1)\pi t_a\big) \op \eta^3}
{\vtgen{\, \frac12}{\frac12 + (2j+1)t_a} } \, 
\prod_{\substack{a=1 \\ mt_a \in \frac12 + \bz}}^{8+d} \frac{\eta^3}
{\vtgen{\, 0}{\frac12 + (2j+1)t_a} } \\[2mm]
& \hspace*{-2mm} \times
e^{i\pi m (2j+1)^2 v^2} \op e^{-i\pi \epsilon j^2 w^2} 
\hspace*{-2mm} \sum_{\substack{P\, \in \, \bz_2\\ \text{classes of}\, I^*}} \hspace*{-2mm} 
\bar q^{\frac12 (P+mv)_R^2} \op e^{-2i\pi (P+mv)\cdot (2j+1)v} \op e^{-i\pi \epsilon (2j+1)  P^2}\ .
 \end{split}\label{zgmgoddfinal}
\ee
The numerical factor in front is given by
\be\label{Fmp}
\cF_m'=2^{n_m-\frac{\ell_2}2}
\ee
for all $j$ excluding $j=\frac{m-1}2$. The overall phase is the combination
\be\label{chiphase_ii}
e^{i\pi\chi_m}:=e^{i\pi\theta_2}e^{i\pi\Delta(m,0)} e^{i\pi \nu(m,0)}e^{i\pi \mu} \ ,\vs{-5}
\ee
where $e^{i\pi\theta_2}$ is the phase of $C_2$ defined in eq. \eqref{c2volmvalue}.

It is crucial to notice that the momentum sum in the above result for $Z(g^m, g^{-(2j+1)})$ is over
$\bz_2$ classes of $I^*$. As shown in eq. \eqref{Im*}, such classes precisely belong to the $g$-invariant subset of $I_m^*$,
which is the momentum lattice in $Z(g^m,\uno)$. Thus, the result fulfills a basic requirement to arise as 
$\text{Tr} g^{-(2j+1)}$ over the Hilbert space in the $g^m$-twisted sector implicit in $Z(g^m,\uno)$. 
We will shortly analyze whether phases and multiplicities are also consistent with operator interpretation.

\subsection{Summary}\label{subsec_sum}

In this section we have determined the full partition function of $\bt^{2d}/\bz_{2m}$ orbifolds by
applying modular transformations to untwisted sector terms and have derived the constraints due
to modular invariance.  In the $g$-twisted sector we have deduced that
$Z(g,\uno)=Z(g,g^{2m})$ requires
\vs{-5}\be\label{lmc_summ}
2mv+w\in I\ ,\qquad 4mE_1+2mv^2\in2\bz .
\ee
Here the shifts $v$ and $w$ are constant vectors along the directions of $\Gamma(16+2d,2d)$ which
are invariant under the $\bz_{2m}$ automorphism. The zero point energy $E_1$ , given in eq. \eqref{ezero1},
depends on the eigenvalues of the automorphism.
In the $g^2$-twisted sector modular invariance reduces to the condition in eq. \eqref{LMCg2_ii} which
involves $w$ and the eigenvalues. Modular invariance in other twisted sectors do not introduce additional
requirements. In the next section we will examine whether there are further constraints arising from the operator
interpretation of the terms in the partition function.

The shift $w$ is intrinsic to the invariant lattice and can be obtained systematically as explained in appendix \ref{app_w}.
It satisfies $w \in I^*$ and $2w \in I$.
Concerning the shift $v$, in principle there can be several solutions to the modular invariance conditions \eqref{lmc_summ}
as discussed in section \ref{section_hm}, where we will also determine additional requisites to avoid massless states in twisted 
sectors. One relevant observation is that when $w$ is non-trivial, the condition $2mv + w \in I$ implies 
$v \notin I^*$ and $mv \notin I^*$, as can be shown using that $I^*/I$ only has $\bz_2$ and $\bz_m$ conjugacy classes.

 \section{Operator interpretation}\label{section_op}

In the preceding section we obtained the full partition function of a class of $\bt^{2d}/\bz_{2m}$ asymmetric orbifolds 
by applying modular transformations to terms in the untwisted sector where the orbifold action $g$ was defined.
In this section we verify validity of the operator interpretation in our asymmetric orbifolds. 
We focus on $\cz_\Gamma(\tau,\bar \tau)$ for the internal bosons, cf. \eqref{fullz}. The task
is to show that $Z(g^\ell, g^k)$ can be understood as $\text{Tr}_{\ch_\ell}\!\! \left(g^k \, q^{L_0} \, \bar q^{\bar L_0}\right)$,
where $\ch_\ell$ is the Hilbert space in the $g^\ell$-twisted sector. 
A basic check is that all $g^k$ act on the same spectrum that appears in $Z(g^\ell,\uno)$. In particular, the momenta that
appear in $Z(g^\ell,g^k)$ must be a subset of the momentum lattice $I^*_\ell$ that occurs in $Z(g^\ell,\uno)$.
For the twisted sectors with $(\ell,2m)=1$ this is satisfied because the lattice sum is over $I^*$ for all $k$, as seen for
instance in eq. \eqref{zggjb}. For the twisted sectors with $\ell$ even and $\ell=m$ we checked it explicitly in 
subsections \ref{subsection_g2} and \ref{subsection_gm} respectively.

Another necessary condition for a valid operator interpretation is that there is a consistent group action. This means that
for any state which appears in both $Z(g^\ell,g)$ and $Z(g^\ell,g^{k})$, the phase of the state in the $(g^\ell,g^{k})$ sector 
must be the $k^{\rm th}$ power of the phase of the state in the $(g^\ell,g)$ sector. Furthermore, the action of $g^k$ on the
degeneracy factor $\cF_{\ell}$ of $Z(g^\ell,\uno)$ must be defined to match the degeneracy factor in $Z(g^\ell,g^{k})$.
When the operator interpretation is valid, the phases and degeneracy factors are such that states surviving the orbifold projection 
via $\frac1{2m}\sum_{k=0}^{2m-1} Z(g^\ell, g^k)$ are guaranteed to have integer multiplicities.

In $g^\ell$-twisted sectors with one orbit, i.e. $(\ell,2m)=1$, both criteria are satisfied. This can be understood by considering the 
$g$-twisted sector partition functions $Z(g,g^j)$ in eq. \eqref{zggjb}, and noticing the $j$-dependence of phases, and that the 
degeneracy factor is $\cF_1$, given in eq. \eqref{Fs}, for all values of $j$. The same result holds for other sectors with a single orbit. 
Thus we need to check consistency of operator interpretation in the remaining twisted sectors. 

Summarising, we will show in detail that consistency of operator interpretation in all twisted sectors imposes the following conditions 
\vs{-5}\be\label{opinttot}
\theta_2+\frac{1-\epsilon}{4}n_1+\frac{\epsilon}2n_m\in2\bz\ ,\qquad
m^{\frac12(n_1-\ell_m^N)}-1\in2\bz_{\ge0}\ ,\qquad 2^{\frac{m-1}2(n_1-\frac{\ell_2^N}{m-1})}-1\in m\bz_{\ge0}\ .
\ee
Remarkably, all these conditions are satisfied for Leech sublattices we use in this work. It would be interesting if they could be 
proven as general lattice properties.

In the untwisted sector we introduced the action of $g$ on $\Gamma$ guided by modular invariance. 
However, we have not proved that vertex operators satisfy consistent operator product expansions. 
As advised in \cite{Harvey:2017rko}, it is then reassuring to verify that untwisted sector states have integer multiplicities.

\subsection{\texorpdfstring{$g^{2j}$}{TEXT}-twisted sectors}\label{subsection_op_g2}

The discussion is split in two parts. We first study the conditions arising from consistency of phases due to the action of
the orbifold generator on momenta and oscillator modes. We then analyze the action on the degeneracy factors.

\subsubsection{Phase}\label{subsection_op_g2ph}

There are 2 $T$-orbits in the $g^{2j}$-twisted sectors. Let us start from the group action on phases in the $g^2$-sector. Consider the 
partition function $Z(g^2,g^2)$ in eq. \eqref{zg2geven}. The action of $g^2$ on momenta in the $g^2$-twisted sector reads
\be
\label{g2g2action}
g^2|P+ 2v + w\rangle = e^{2\pi i E_2}\,  e^{i\pi (P + 2v + w)^2}|P+ 2v + w\rangle \ ,\vs{-5}
\ee
for $P \in I_2^*$. Next consider the $(g^2,g^{-1})$ orbit. From eq. \eqref{Zg2gm1}, we obtain that in the $g^2$ sector the action 
of $g^{-1}$ on momenta is given by
\vs{-5}\bea\label{g2gm1action}
&&g^{-1}|P + 2b_i f_i +2v + w\rangle = \\
&&\qquad\qquad e^{\frac{i\pi d}{2}}  e^{i \pi \theta_m^1}  e^{-i\pi  E_2}\,  e^{2\pi iv^2} 
\, e^{-2i\pi (P+2 b_i f_i + 2v+w)\cdot v} \, e^{-2\pi i(b_i f_i)^2}|P + 2b_i f_i +2v + w\rangle\ ,\nn
\eea
for $P \in I$. By consistency, the square of the right hand side of the above times the right hand side of eq. \eqref{g2g2action} must be 
equal to 1, i.e.
\vs{-5}\be\label{checkg2a}
e^{i \pi d}\op  e^{2 \pi i\theta_m^1} \op e^{4\pi iv^2} \op e^{-4i\pi (P+2 b_i f_i + 2v+w)\cdot v} \op 
e^{-4\pi i(b_i f_i)^2} \op e^{i\pi (\hat P + 2v + w)^2} =1\ ,\vs{-5}
\ee
for all $P$ in $I$. Here $\hat P  \in I_2^*$ and it must be $\hat P = (P + 2b_i f_i)$ because these are the vectors of $I_2^*$ which are 
invariant under $g$. To simplify the left hand side notice that $\hat P$ belongs to a $\bz_m$ class of $I^*$, and recall also that $w$ is in 
a $\bz_2$ class of $I^*$. Therefore $\hat P \cdot w \in \bz$ because $m \hat P \cdot w \in \bz$, $\hat P \cdot 2w \in \bz$, and $m$ is 
odd. Moreover
\vs{-5}\be 
\hat P^2 = P^2 + 2P \cdot (2b_i f_i) + (2b_i f_i)^2 = 4 (b_i f_i)^2 \ \text{mod} \ 2  \quad\forall\;P\in I\ .\vs{-5}
\ee
Eq. \eqref{checkg2a} then reduces to
\vs{-5}\be
\label{checkg2b}
w^2 + 2 \theta_m^1 + d \in 2\bz \ .
\ee
In appendix \ref{appC2} we prove that this condition always holds for the lattice $I^*$. Following the same steps one may verify 
consistency of the $g^{2j-1}$-action in the $g^2$ sector.

A similar analysis for $g^{2j}$-twisted sectors, $0<j<m$, gives the condition 
\vs{-5}\be\label{check2c}
j^2w^2 + 2j \theta_m^j + jd \in 2\bz \, .
\ee 
However, from properties of the $\theta_m^j$ shown in appendix \ref{appCma}
and eq. \eqref{checkg2b}, it follows that this requirement does not lead to additional constraints. 
For $j=2l$ it is always satisfied noting that $2w\in I$ and that 
$4\theta_m^{2l}\in2\bz$, cf. \eqref{them1phi}. For odd values $j=(2l+1)$ we also need to use  
that $2\theta_m^{2l+1} = 2 \theta_m^1 \, \text{mod} \, 2$ and $4\theta_m^1 \in 2\bz$.

\subsubsection{Action on degeneracy factors}\label{subsection_op_g2mult}

We now consider the modulus of the overall constants that appears in $Z(g^2,g^l)$, $0\le l<2m$. These numbers have the interpretation 
of traces of $g$ to various powers over the set of $\cF_2$  degenerate states, where $\cF_2$ is given in eq. \eqref{Fs}.
In the $(g^2,\uno)$ sector, this constant is just Tr of $\uno$ over the degeneracy factor, which gives $\cF_2$. Then in the $(g^2,g)$ sector, 
the overall constant will be ${\rm Tr}\,(g)$ over $\cF_2$ degenerate states. Since the order of $g$ is $2m$, the eigenvalues of the
$g$-action on $\cF_2$ will in general be some $\omega$, with $\omega^{2m}=1$. However, there is a constraint. Since in $(g^2,g^2)$ 
sector the modulus of the overall constant is again $\cF_2$ (as can be seen from $T$-transformation), then $g^2$ on all the $\cF_2$ 
states must give the same phase, say $\omega^2$. This means that $g$ must act as $\pm\omega$ on these degenerate states. 
Suppose there are $\cF_2^+$ eigenvalues that are $+\omega$, and $\cF_2^-= \cF_2-\cF_2^+$ eigenvalues that are $-\omega$. 
Then ${\rm Tr}\,(g)$ over the $\cF_2$ states will give $(\cF_2^+-\cF_2^-)\omega$. We can continue this argument for 
other $(g^2,g^{2k+1})$ sectors ($0\!\le\! k\!\le\! m-1$) and the result is $(\cF_2^++(-1)^k \cF_2^-) \omega^{2k+1}$. 
Thus, by consistency, the modulus of the constant appearing in $Z(g^2,g)$, which we denoted as $\cF_2'$ in eq. \eqref{Zg2gm1}, 
must be $\cF_2$ minus some non-negative even number (this is because $\cF_2^+-\cF_2^-= \cF_2- 2 \cF_2^-$). 

From eqs. \eqref{Fs} and \eqref{F2p} we find
\vs{-5}\be
\mathcal F_2-\mathcal F_2'=m^{\frac12(n_2-\ell_m)}(m^{\frac12(n_1-\ell_m^N)}-1)\ .\vs{-5}
\ee
Since $n_2-\ell_m\in2\bz_{\ge0}$ -- cf. eq. \eqref{Fint} -- and $m$ is odd, the consistency condition
$\mathcal F_2-\mathcal F_2' \in 2\bz_{\ge0}$ requires
\vs{-5}\be\label{opintg2fx}
m^{\frac12(n_1-\ell_m^N)}-1\in2\bz_{\ge0}\ ,\vs{-5}
\ee
i.e. $n_1-\ell_m^N\in2\bz_{\ge0}$. The same result holds for other $g^{2j}$-twisted sectors, $0<j<m$.

\subsection{\texorpdfstring{$g^{m}$}{TEXT}-twisted sector}\label{subsection_op_gm}

We again discuss separately consistency of phases and proper action on degeneracy factors.

\subsubsection{Phase}\label{subsection_op_gmph}

The $g^m$-twisted sector has $m$ orbits. For the $(g^m,\uno)$ orbit, from eq. \eqref{zgmgma}, we deduce that the action of $g^m$ on momenta is
\be
\label{gmgmaction}
g^m|P+ mv\rangle = e^{2\pi iE_m} e^{-i\pi  m^2 v^2 }\, e^{2\pi i(P + mv) \cdot mv}\, e^{i\pi P^2}|P+ mv\rangle\ ,\vs{-5}
\ee
for $P \in I_m^*$. For other orbits $(g^m,g^{-(2j+1)})$, from eq. \eqref{zgmgoddfinal}, we notice that the momenta are shifted by $mv$ as in $Z(g^m,\uno)$. Moreover, the insertion of $g^{-(2j+1)}$ forces the momenta to be in uncorrelated $\bz_2$ classes of $I^*$, which precisely form the \mbox{$g$-invariant} part of $I_m^*$. We thus find that the action of $g^{-(2j+1)}$ on these momenta is
\be
\label{gmgoddaction}
g^{-(2j+1)}|P+ mv\rangle =e^{i\pi \chi_m} e^{i\pi m(2j+1)^2 v^2} \op e^{-i\pi \epsilon j^2 w^2} \op
e^{-2\pi i(P + mv)\cdot (2j+1)v}\op e^{-i\pi \epsilon (2j+1) P^2}|P+ mv\rangle\ .\vs{-5}
\ee
We are now ready to check the operator interpretation.

Setting $j=0$ in eq. \eqref{gmgoddaction} shows that the $g^{-1}$-action in the $g^m$-sector is
given by
\be
\label{gmg1action}
g^{-1}|P+ mv\rangle =e^{i\pi  \chi_m} e^{i\pi m v^2} \op e^{-2i\pi (P + mv)\cdot v}\op e^{-i\pi \epsilon P^2}|P+ mv\rangle\ .\vs{-5}
\ee
Consistency of the group action requires that the \mbox{$m^{\rm th}$} power of the above right hand phase times the phase in  the right hand side of $g^m|P + mv\rangle$ in eq. \eqref{gmgmaction}, must be equal to one, i.e.
\be
\left[ e^{i\pi \chi_m} e^{i\pi m v^2} \op e^{-2i\pi (P + mv)\cdot v}\op e^{-i\pi \epsilon P^2}\right]^m
\left[ e^{2\pi iE_m} e^{-i\pi m^2 v^2} \op e^{2i\pi (P + mv)\cdot mv}\op e^{i\pi P^2}\right] = 1 \, ,
\ee
for all $P$ in $\bz_2$ classes of $I^*$. Simplifying gives
\be
e^{i\pi m\chi_m} e^{2\pi iE_m} = 1 \ ,
\ee
where we used that $e^{i\pi (1- \epsilon m) P^2} = 1$, since $m=\epsilon \ \text{mod} \ 4$, $\epsilon^2=1$, and 
$4P^2=\text{even}$ for $P$ in $\bz_2$ classes of $I^*$. 
Substituting eqs. \eqref{ezerom} and \eqref{chiphase_ii}, and using \eqref{delnumu}, then implies that consistency requires
\vs{-5}\be\label{opent_gm_ph}
\theta_2+\frac{1-\epsilon}{4}n_1+\frac{\epsilon}2n_m\in2\bz\ .\vspace{-5pt}
\ee
Notice that this constraints $2\theta_2$ to be integer.
We remark that eq. \eqref{opent_gm_ph} is stronger than the level matching condition in the even-twisted sectors, 
given in eq.\eqref{LMCg2_ii}. Indeed, the latter can be written as
\vs{-5}\be\label{LMCg2_iii}
2\theta_2+\frac{1-\epsilon}{2}n_1+\epsilon n_m\in2\bz \, ,
\ee
by virtue of the relation $w^2=2\epsilon \theta_2  \, \text{mod}\, 2$, proven in appendix \ref{appC2}. 

Consistency of the operator interpretation also requires that the $(2j+1)^{\rm th}$ power of the phase in $g^{-1}|P + mv\rangle$ action, c.f. eq. \eqref{gmg1action}, must be equal to the phase in the $g^{-(2j+1)}|P + mv\rangle$ action, c.f. eq. \eqref{gmgoddaction}. 
This condition is satisfied provided that
\be
e^{2 \pi i\,j\chi_m}\op e^{-2 \pi i\,j(2j+1) m v^2} \op e^{i\pi \epsilon j^2 w^2} = 1\ .
\ee
We may use the relation $w^2=2\epsilon \theta_2  \, \text{mod}\, 2$, cf. eq. \eqref{w2th2}, and the $g$-twisted sector level matching 
condition \eqref{LMCg} together with eq. \eqref{ezero1}, to express $w^2$ and $v^2$ in terms of eigenvalue multiplicities. The above 
condition then reduces to  $2\theta_2\in\bz$. This is however not a new constraint, as it is already required to hold by the 
consistency condition \eqref{opent_gm_ph}.

\subsubsection{Action on degeneracy factors}
\label{subsection_op_gmmult}

The same argument used in subsection \ref{subsection_op_g2mult} now asserts that if $\cF_m$ and $\cF_m'$ are the degeneracy factors 
in the $(g^m,\uno)$ and $(g^m,g)$ sectors, then $\cF_m-\cF_m'$ must be a multiple of $m$, i.e.
\be\label{opintgm}
\cF_m-\cF_m'=\mathfrak{b} m\vs{-5}
\ee
for some non-negative integer $\mathfrak{b}$. Then $g$ has a cyclic $\bz_m$ action on $\mathfrak{b}$ copies of $m$-plet states, 
and these $\mathfrak{b} m$ states drop out of the ${\rm Tr}(g)$ over $\cF_m$ states. The remaining number will then be 
$\cF_m-\mathfrak{b} m=\cF_m'$. The modulus of the overall constants that appear in $Z(g^2,g^{2j+1})$ sectors, for 
$0\le j<m$ and $2j+1\ne m$, are all the same, and so the same result holds for action of $g^{2j+1}$. 

From eqs. \eqref{Fs} and \eqref{Fmp} we find
\vs{-5}\be\label{opintgmfxa}
\mathcal F_m-\mathcal F_m'=2^{n_m-\frac{\ell_2}2}\Big(2^{\frac12((m-1)n_1-\ell_2^N)}-1\Big)\ .\vs{-5}
\ee
The condition \eqref{opintgm}, with $\mathfrak{b}\in\bz_{\ge0}$, is then satisfied provided
\be\label{opintgmfx}
2^{\frac{m-1}2(n_1-\frac{\ell_2^N}{m-1})}-1\in m\,\bz_{\ge0}\ .
\ee
The possible odd prime numbers $m$ which appear in $\bz_{2m}$ automorphisms of the Leech lattice are $m=\{3,5,7,11,13,23\}$ 
(they divide the order of the automorphism group of the lattice, i.e. the Conway group Co$_0$). It turns out that the above condition is valid if
\be\label{opintgmfxb}
n_1-\frac1{m-1}\,\ell_2^N\in2\bz_{\ge0}\ ,\quad m\ne\{7,23\}\ ,\qquad\qquad n_1-\frac1{m-1}\,\ell_2^N\in\bz_{\ge0}\ ,\quad m=\{7,23\}\ .\vs{-5}
\ee
Note that this means $\ell_2^N\in(m-1)\bz_{\ge0}$.

\subsection{Untwisted sector}\label{subsection_op_un}

In $Z(\uno, g^k)$, the action of $g$ on oscillator modes is captured in $\vartheta$ functions and is clearly consistent with
the group law. On the other hand, the action on momenta manifests in phases and lattice sums that depend on $k$. 
Then, in the following we only pay attention to the action on momenta. The phases depend on the constant shifts
$v$ and $w$ which satisfy $2mv + w \in I$, $w \in I^*$ and $2w \in I$.

In the beginning there is $Z(\uno,\uno)$ in which the momentum lattice is the full $\Gamma$. It helps to recall that a generic 
$P \in \Gamma$ has components $(P_I,P_N)$ with $P_I \in I^*$ and $P_N \in N^*$. More precisely, as shown in eq. \eqref{Gamma},
$\Gamma$ can be decomposed as the direct sum of $(I,N)$ plus correlated $\bz_2$ classes and $\bz_m$ classes.

Let us first consider momenta with $P_N=0$, which means $P \in I$. Since $I$ is a subset of $I_2$ and $I_m$, states
with $P \in I$ will appear in all $Z(\uno,g^k)$, $k=0, \ldots, (2m-1)$, and with a phase $e^{2 i\pi P\cdot kv}$.
Here $w$ drops out because $P\cdot w \in \bz$ for $P \in I$. Notice also that $\big(e^{2i\pi P\cdot v})^{2m} =1$, since 
$2mv = w \, \text{mod}\, I$. But then $\frac1{2m} \sum_{k=0}^{2m-1} e^{2 i\pi P\cdot kv}$ equals 0 unless
$e^{2i\pi P\cdot v}=1$. Therefore, only states with $P\cdot v \in \bz$ survive the orbifold projection and do so with multiplicity 1.

Next we check states with $P_I =0$, so that $P \in N$. Since $v$ and $w$ only have components along 
invariant directions, these states will not pick up momentum dependent phases under $g$.
There are 3 possibilities. First take $P$ along invariant directions of $\Theta^2$, i.e. $P \in N_2$ which also occurs
in the invariant sublattice $I_2$, cf. \eqref{I2Im}. Then such $P \in N_2$ will appear in $Z(\uno,g^{2j})$,
$j=0,\cdots,(m-1)$, and the orbifold projection gives a factor $\frac12$. But there are precisely 2 elements
in the orbit,  since $\Theta^2 P =P$ for $P \in N_2$, and we can form the invariant combination
$|P\rangle + |-P\rangle$. To continue take $P$ such that $\Theta^m P=P$, so that $P \in N_m \subset I_m$, as
seen in \eqref{I2Im}. Thus, $P \in N_m$ enters in $Z(\uno, \uno)$ and $Z(\uno,g^m)$,
the orbifold projection gives a factor $\frac1{m}$, as expected for an orbit of length $m$, and there is an
invariant  combination $\sum_{j=1}^{m-1} \Theta^j |P \rangle$. The third case is when $P\in N$ is in a
generic orbit with $2m$ elements. These states only appear in $Z(\uno, \uno)$ and we can form an invariant combination 
$\sum_{k=0}^{2m-1} \Theta^k |P\rangle$.

Finally we look at states with $P_I \ne 0$ and $P_N \ne 0$. Since we already treated the cases $P \in I$ and $P \in N$, 
we will restrict to momenta that belong to the correlated classes to simplify the discussion.
The decomposition of $\Gamma$ in eq. \eqref{Gamma} indicates that there are 3 cases that we now analyze in turn.
\vskip 0.1in
\noindent{\bf Case 1.} $P$ in $\bz_2$ correlated  classes, i.e. $P= \sum\limits_{i=1}^{\ell_2} a_i (f^I_{2,i}, f^N_{2,i})$.
\vskip 0.1in
\noindent
Notice that these $P$ belong to $I_2$, as shown in eq. \eqref{I2Im}. Thus, these states will appear in
all $Z(\uno,g^{2j})$, $j=0,\ldots, (m-1)$, with phase $e^{2 i\pi j P\cdot (2v+w)}$, using $2w \in I$.
But $e^{2 i\pi m P\cdot (2v+w)}=1$, since $m(2v +w) \in I$ by assumption. Hence, 
$\frac1{2m} \sum_{j=0}^{m-1} e^{2 i\pi j P\cdot (2v +w)}$ vanishes unless
$e^{2i\pi P\cdot(2 v +w)}=1$. Thus, only states with $P\cdot(2 v+w) \in \bz$ are not projected out.
There will be a factor $\frac12$ which just means that only a symmetric combination of the states
$| \sum_i a_i (f^I_{2,i}, f^N_{2,i})\rangle$ and $|\sum_i a_i (f^I_{2,i},- f^N_{2,i})\rangle$ actually survives.
Recall that $\Theta^2  f^N_{2,i} = f^N_{2,i}$ and $\Theta  f^N_{2,i} = -f^N_{2,i}$.

\vskip 0.1in
\noindent{\bf Case 2.} $P$ in $\bz_m$ correlated classes, i.e. $P= \sum\limits_{j=1}^{\ell_m} b_j (f^I_{m,j}, f^N_{m,j})$.
\vskip 0.1in
\noindent
According to eq. \eqref{I2Im}, these momenta also belong to $I_m$ and the corresponding states arise
in $Z(\uno,g^{lm})$, $l=0, 1$. Now $g$ acts with phase $e^{2 i\pi l P\cdot m v}$ and
$e^{2 i\pi P\cdot m v}$ is clearly $\pm 1$. Summing over $l$ then gives that only states with $P\cdot m v \in \bz$
survive with a factor $\frac1{m}$. But in this case the orbit of $\Theta$ has precisely $m$ elements, since
$\Theta^m  f^N_{m,j} = f^N_{m,j}$.
Thus, the fractional factor means that only the invariant combination $\sum_{n=0}^{m-1}
|\sum_j b_j (f^I_{m,j}, \Theta^n f^N_{m,j}) \rangle$ will survive.

\vskip 0.1in
\noindent{\bf Case 3.} $P$ in a combination of $\bz_2$ and $\bz_m$ correlated classes.
\vskip 0.1in
\noindent
States with these momenta will only show up in $Z(\uno,\uno)$. The factor $\frac1{2m}$ in the orbifold
projection is due to the fact that the orbit of $P$ has length $2m$ and instructs us to form the
invariant combination in the obvious way.

In conclusion, we have verified that all multiplicities in the untwisted sector are positive integers.

\section{Examples}\label{section_hm}
In this section we construct heterotic asymmetric orbifolds with rank reduction, including islands, in the framework developed in
the preceding sections.
Each model is identified as HM\#, where \# is the entry number in Table 1 of \cite{HM}.
The normal lattice $N$ is taken to be the coinvariant lattice corresponding to the entry.
We will specify the invariant lattice $I$ which we glue with $N$ to construct $\tdl$. 
The automorphism $\Theta$ of order $2m$, with $m$ prime different from 2, is obtained
as explained in subsection \ref{subsection_leech}. 
For each model, we find the characteristic vector $w$ associated to $I^*$ and then
determine shift vectors $v$ that satisfy the modular invariance conditions summarized in eq. \eqref{lmc_summ}.

In all examples the lattice $I$ is purely right-moving, i.e. with signature $(0,2d)$, whereas $N$ is purely left,
with signature $(16+2d,0)$.
In appendix \ref{apprr} we describe two models with rank reduced gauge groups and with indefinite invariant lattices.

The spectrum of the theory is obtained from the full partition function given in eqs. \eqref{genpf}-\eqref{fullz}. 
We discuss first the untwisted sector. The masses of states, read from eq. \eqref{z1gl}, are given by
\be\label{massunt}
m_L^2=\frac12P_L^2+N_L-1\ ,\qquad\qquad m_R^2=\frac12P_R^2+N_R+\frac12r^2-\frac12\ ,
\ee
where $N_L$ and $N_R$ are left- and right-moving oscillator numbers, $r^2\ge1$ and $r$ is in vector or spinor classes of $SO(8)$. 
A massless state has $m_L^2=m_R^2=0$. In the right-moving sector this requires $P_R=0$, $N_R=0$ and $r^2=1$. 
In the left-moving sector, either $P_L^2=2$ or $N_L=1$ must hold. Since the normal lattice $N$ is purely left and is a sublattice 
of the Leech lattice, it has no root vectors, i.e. $P_L^2\ne2$. It is however possible to have $N_L=1$. The oscillators along spacetime 
directions are $g$-invariant and combine with right-moving $g$-invariant modes to give the gravity multiplet. The $16+d$ oscillators 
along $N$-directions are projected out under the orbifold action and do not give massless states. Thus there are no vector multiplets in the untwisted sector.

In the  twisted sectors the spectrum follows from eqs. \eqref{zggjb}, \eqref{zgevengeven}, \eqref{zgevengodd}, \eqref{zgm1}and \eqref{zgmgoddfinal}. The masses are determined by
\be\label{masstw}
m_L^2=\frac12P_L^2+N_L+E_\ell-1\ ,\qquad\qquad m_R^2=\frac12(P+\ell v+w\delta_{\ell,2\!\!\!\!\!\mod\!4})_R^2+N_R+\frac12r^2-\frac12\ ,
\ee
where $E_\ell$ is the zero point energy in the $g^\ell$-twisted sector and $P$ is in $I_\ell^*$. It is shown in appendix \ref{app_w} that 
the shift vector $w$ is non-trivial only if there exist vectors $P\in\bz_2$ classes of $I^*$ such that $P^2$ is $1/2$ mod integers. 

Let us first consider  theories with non-trivial $w$. In this case $2mv=w\, \text{mod}\,  I$. In $g^\ell$-twisted sectors with $(\ell,2m)=1$, 
the right-moving momentum is $P+\ell v$, where $P\in I^*$. Since $\ell v\notin I^*$, $P+\ell v\ne 0$ and so there are no 
massless states in these sectors. In the $g^m$ sector, the right-moving momentum is $P+m v$, where $P$ is in $\bz_2$ classes of 
$I^*$. Since $mv$ is not in $I^*$, there are no massless states in the $g^m$-twisted sector either. In $g^{2j}$-twisted 
sectors ($0<j<m$), the right-moving momentum is $P+2jv+w\delta_{2j,2\!\!\!\mod\!4}$, where $P$ is in $\bz_m$ classes of $I^*$. 
Therefore, to avoid massless states in even-twisted sectors, shift vectors must satisfy
$2jv+w\delta_{2j,2\!\!\!\mod\!4}\notin\bz_m$ classes of $I^*$. Since $2mv + w \in I$, $w \in I^*$, and $2w \in I$, this
condition simplifies to $2v \notin I^*$. 

All in all, when $w$ is non-trivial and $2 v \notin I^*$, the resulting theory has massless states only in 
the gravity multiplet in the untwisted sector, and thus constitutes a heterotic island. We construct two island models in 
subsections \ref{section_hm149} and \ref{section_hm251}. 
These are $\bz_{10}$ and $\bz_{22}$ orbifolds in respectively 6 and 4 dimensions. They correspond to Leech sublattices 
HM149 and HM251. In fact, for each model, there might be non-equivalent shift vectors $v$, 
satisfying all modular invariance conditions and $2 v \notin I^*$, for which the massive spectra of the orbifolds are different, and so lead
to different islands. As explained in appendix \ref{appnoneqv}, by non-equivalent shift vectors we mean vectors $v_1$ and $v_2$ such that
\be\label{vequiv}
zv_1-\theta v_2\notin I^*, \, \qquad (z,2m)=1, \,\qquad z\in \mathbb{Z}\, ,
\ee
for all point automorphisms $\theta$ of $I$ (i.e. automorphisms which fix the origin and leave the Gram matrix of $I$ invariant).  
When there are inequivalent shifts there will rather be an archipelago instead of an island.

We next consider theories which have a trivial shift vector $w$, i.e. $P^2\in\bz$ $\forall P\in\bz_2$ classes of $I^*$. In this case 
$2mv\in I$. Following the above arguments, there are no vector multiplets in the untwisted sector. If $4mE_1$ is even, then $v=0$ 
is a solution to the level matching condition $4mE_1+2mv^2\in2\bz$, and with no shift there will be massless states in all twisted 
sectors. For appropriate choices of $v\ne 0$ one can eliminate some of the massless states in twisted sectors. The resulting theory is therefore 
not an island, but can have gauge group of reduced rank. Model HM100 in subsection \ref{section_hm100} provides such an example.

Other potential rank-reduced (but not island) $\bt^{2d}/\bz_{2m}$ orbifolds, which we do not discuss further, correspond to 
HM122 ($\bz_{10}$), HM129 ($\bz_{14}$), HM159 ($\bz_{10}$) and HM232 ($\bz_{14}$).

\subsection{\texorpdfstring{$6d$}{TEXT} Islands HM149}\label{section_hm149}

The first example is a $\bz_{10}$ orbifold in 6 dimensions, i.e. the normal lattice $N$ and invariant lattice $I$ have respectively 
signature $(20,0)$ and $(0,4)$. The eigenvalue distribution \eqref{ta} of $\Theta$ is $t=\tfrac1{10}(1,1,2,2,3,3,4,4,5,5)$, which 
means \mbox{$n_1=n_2=n_5=2$}, which also follow from the frame shape $2^2\cdot 10^2$ \cite{HL}.
The zero point energies are then found to be
\be\label{hm149Es}
E_1=\frac{19}{20}\ ,\qquad\qquad E_2=\frac{4}{5}\ ,\qquad\qquad E_5=\frac{3}{4}\ .\vs{-5}
\ee

Consider the normal lattice $N$ and the invariant lattice $\tilde I$ of entry HM149 in \cite[Table 1]{HM}. 
The normal lattice has discriminant group $N^*/N \cong \bz_2^2\times\bz^2_{10}$.
To construct the even self-dual lattice $\Gamma(20,4)$,  we may glue the lattice $N$ with either $I=\tilde I(-1)$, i.e. $\tilde I$ with 
reversed signature, or with the orthogonal complement of $\tilde I$ in the root lattice of $E_8$, denoted by $\tilde I'$, with reversed 
signature. Reference \cite{Baykara:2021ger} provides 
a Mathematica package which, among other things, finds a primitive embedding of all $\tilde I$ with dimension smaller than 5 in 
$E_8$, and computes their orthogonal complements $\tilde I'$. We have used this package to compute the Gram matrix of $\tilde I'$. 
The Gram matrices are
\be\label{I149}
G_{\tilde I}={\scriptsize \begin{pmatrix}
4&0&0&2\\
0&4&2&0\\
0&2&6&0\\
2&0&0&6
\end{pmatrix}}\ ,\qquad\qquad\qquad
G_{\tilde I'}={\rm diag~}(2,2,10,10)\ .\vs{-5}
\ee
The two lattices $\tilde I$ and $\tilde I'$ are not isometric. Notice also that $\tilde I' = 2 A_1 + 2 A_1(5)$.
Besides, it can be checked that $\tilde I$ is isomorphic to 
$\big(\begin{smallmatrix}
6 & 4 \\
4 & 6
\end{smallmatrix}\big)^{\!\oplus\op 2}$. 

The discriminant groups of invariant lattices $I$, $I_2$ and $I_5$ are given by
\be\label{I*149}
I^*/I \cong \bz_2^2\times\bz_{10}^2 \cong \bz_2^4\times\bz_{5}^2\ ,
\qquad\qquad I_2^*/I_2\cong\bz_5^4\ ,\qquad\qquad I_5^*/I_5\cong\bz_2^{12}\ .
\ee
From eq. \eqref{discIs}, we find $\ell_2=4$, $\ell_2^N=8$, $\ell_5=2$ and $\ell_5^N=2$. The degeneracy factors, given in eqs. \eqref{Fs}, \eqref{F2p} and \eqref{Fmp}, are then found to be
\be\label{F149}
\cF_1=1\ ,\qquad\qquad \cF_2=\cF_2'=1\ ,\qquad\qquad \cF_5=\cF_5'=1\ .\vs{-5}
\ee
The quantities $C_2$ \eqref{C2} and $C_5(1)$ \eqref{Cma} have values
\be\label{C5C2hm149}
C_2=-4\ ,\qquad\qquad\qquad C_5(1)=5\ ,\vs{-5}
\ee
which, using eqs. \eqref{c2volmvalue} and \eqref{Cma2}, yield $\theta_2=1$ and $\theta_{5}^1=0$. Notice that while we have proved 
in appendix \ref{appCma} that $|C_2|=\sqrt{2^{\ell_2}}$ and $|C_m(a)|=\sqrt{m^{\ell_m}}$, we do not have general formulae for 
the phases $\theta_2$ and $\theta_m^1$, but rather compute them case by case. We observe that the consistency conditions of 
operator interpretation \eqref{opinttot} are all satisfied. 

The dual lattice $I^*$ has $\bz_2$ classes with half-integer length squared and thus admits a non-trivial shift vector $w$
that can be found as explained in appendix \ref{app_w}. For $I=\tilde I'(-1)$ with the diagonal Gram matrix given in eq. \eqref{I149}, 
we find that $w=\frac12(1,1,1,1)$ in the basis of $I$. 
For $I=\tilde I(-1)$, the result is $w=\frac12(1,1,0,0)$.
In both cases $w^2$ is even and the modular invariance condition \eqref{LMCg2} is satisfied.

We next determine the shift vectors $v$ that satisfy the modular invariance conditions \eqref{lmc_summ}, as well as the condition 
$2 v \notin I^*$ to avoid massless states in twisted sectors. 
For both choices of invariant lattice we find only one possible shift up to equivalences.
The unique solution can be taken to be  $v=\frac1{20}(1,3,1,1)$  for $I=\tilde I'(-1)$ and $v=\frac1{20}(3,1,0,0)$ 
for $I=\tilde I(-1)$.

An important question is whether the models obtained from the two non-isometric invariant lattices $I$ and $I'$ are actually equivalent. 
It is possible that this is the case, since in \cite{Fraiman:2022aik} it is argued, based on the uniqueness of the charge lattice genus, 
that there is only one $\bt^4/\bz_{10}$ heterotic island. 
However, we have verified, by computing the expansion of the corresponding partition functions in powers of  $q\bar q$,  
that the massive spectra  do not agree and, therefore, there are two six dimensional heterotic islands.  
As indicated in \cite{Baykara:2025lhl},
uniqueness can be recovered by adding an extra circle to obtain a compactification to five dimensions. In this case the two invariant 
lattices become indefinite of signature $(1,5)$ and the two models are connected by a $SO(1,5)$ T-duality transformation. This follows 
from the fact that the two indefinite invariant lattices have the same genus and the determinant is such that there is 
only one isometry class -see Corollary 22 in Chapter 15 of \cite{Conway98}.

\subsection{\texorpdfstring{$4d$}{TEXT} Archipelago HM251}\label{section_hm251}

The second example is a $\bz_{22}$ orbifold in 4 dimensions, where the signature of $N$ and $I$ is respectively $(22,0)$ and $(0,6)$. 
The eigenvalues of $\Theta$ have distribution 
$t=\tfrac1{22}(1,2,3,4,5,6,7,8,9,10,11)$, i.e. \mbox{$n_1\!=\!n_2\!=\!n_{11}\!=\!1$}, in agreement with the
frame shape $2\cdot 22$ \cite{HL}.  The zero point energies are
\be\label{hm251Es}
E_1=\frac{43}{44}\ ,\qquad\qquad E_2=\frac{10}{11}\ ,\qquad\qquad E_{11}=\frac{3}{4}\ .
\ee
The normal lattice has discriminant group $N^*/N \cong \bz_2\times\bz_{22}$.
The invariant lattice $I$ is chosen to be the orthogonal complement of $\tilde I$ in the root lattice of $E_8$ with reversed signature, 
and its Gram matrix is given by 
\be\label{I251}
G_{I}=-{\scriptsize \begin{pmatrix}
2 & 0 & 0 & 0 & 0 & 0 \\
 0 & 2 & 0 & 0 & 0 & 1 \\
 0 & 0 & 2 & 1 & 1 & 1 \\
 0 & 0 & 1 & 2 & 0 & 1 \\
 0 & 0 & 1 & 0 & 2 & 0 \\
 0 & 1 & 1 & 1 & 0 & 4 \\
\end{pmatrix}}\ ,
\ee
which was obtained using the package provided in \cite{Baykara:2021ger}.

The discriminant groups of invariant lattices are
\be\label{I*251}
I^*/I \cong \bz_2\times\bz_{22}\cong \bz_2^2\times\bz_{11}\ ,\qquad\qquad 
I_2^*/I_2\cong\bz_{11}^2\ ,\qquad\qquad I_{11}^*/I_{11}\cong\bz_2^{12}\ ,\vs{-5}
\ee
which lead to $\ell_2=2$, $\ell_2^N=10$, $\ell_{11}=1$ and $\ell_{11}^N=1$. The degeneracy factors are
\be\label{F251}
\cF_1=1\ ,\qquad\qquad \cF_2=\cF_2'=1\ ,\qquad\qquad \cF_{11}=\cF_{11}'=1\ ,\vs{-5}
\ee
and the values of $C_2$ and $C_{11}(1)$ are
\be\label{C5C2hm251}
C_2=2\ ,\qquad\qquad\qquad C_{11}(1)=-i\sqrt{11}\ ,\vs{-5}
\ee
i.e. $\theta_2=0$ and $\theta_{11}^1=-\frac12$. Consistency conditions \eqref{opinttot} are again all satisfied. 

For the characteristic vector we obtain \mbox{$w=\frac12(1,1,0,1,1,0)$} in the basis of $I$. 
For shift vectors that satisfy conditions \eqref{lmc_summ} and $2 v \notin I^*$, we find that
there are 232 non-equivalent solutions leading to theories without vector multiplets.
We have therefore located a populated $\bz_{22}$ archipelago in 4 dimensions, formed by islands corresponding to inequivalent shifts.
A natural question is whether the massive spectra of the islands is the same or not.
Since now the $I$ lattice is the same, the question is whether two inequivalent
shifts $v_1$ and $v_2$, i.e. satisfying condition \eqref{vequiv}, give the same
spectrum or not. As argued in appendix \ref{appnoneqv}, in this case it follows that $v_1$ and $v_2$ will lead to different
projections in the untwisted sector and also to different massive states in each sector.
As a further check we have computed the power expansions in $q\bar q$ of the partition function corresponding to 
some inequivalent shifts and found that they are indeed different.

Other $4d$ string vacua with 16 supercharges and no vector multiplets are known in the literature.
In \cite{Dabholkar:1998kv} a $4d$ island was constructed as a $\bt^6/\bz_9$ asymmetric orbifold of type II.
It would be interesting to check whether the shift necessary for modular invariance is unique up to isometries of
the underlying $(0,6)$ invariant lattice. 
On the other hand, a heterotic island was identified in \cite{Chaudhuri:1995fk} using the fermionic formulation.

\subsection{\texorpdfstring{$6d$}{TEXT} rank-reduced HM100}\label{section_hm100}

The last example is a $\bz_{10}$ orbifold in 6 dimensions, i.e. $N$ and $I$ have signatures $(20,0)$ and $(0,4)$. 
The automorphism $\Theta$ has eigenvalues determined by $t=\tfrac1{22}(1,1,1,2,3,3,3,4,5,5)$, i.e. $n_1=3$, $n_2=1$ and $n_{5}=2$,
which are also obtained from the frame shape $1^2 \cdot 2 \cdot 10^3/5$ \cite {HL}.
The zero point energies are
\be\label{hm100Es}
E_1=\frac{9}{10}\ ,\qquad\qquad E_2=\frac{4}{5}\ ,\qquad\qquad E_{5}=1\ .\vs{-5}
\ee
The normal lattice has discriminant group $N^*/N \cong \bz_2^3\times\bz_{10}$.
As for the invariant lattice, we may again choose $I=\tilde I(-1)$ or $\tilde I'(-1)$. The Gram matrices are 
\be\label{I100}
G_{\tilde I}={\scriptsize \begin{pmatrix}
 4 & -2 & -2 & 2 \\
 -2 & 4 & 0 & 0 \\
 -2 & 0 & 4 & -2 \\
 2 & 0 & -2 & 4 \\
\end{pmatrix}}\ ,\qquad\qquad\qquad
G_{\tilde I'}={\scriptsize \begin{pmatrix}
 4 & 2 & 2 & 2 \\
 2 & 4 & 2 & 2 \\
 2 & 2 & 4 & 2 \\
 2 & 2 & 2 & 4 \\
\end{pmatrix}}\ .
\ee
The two lattices are isometric. In fact, $\tilde I \cong \tilde I' \cong A_4(2)$.

The discriminant groups of invariant sublattices are
\be\label{I*100}
I^*/I \cong \bz_2^3\times\bz_{10} \cong \bz_2^4\times\bz_{5}\ ,\qquad\qquad 
I_2^*/I_2\cong\bz_5^4\ ,\qquad\qquad I_5^*/I_5\cong\bz_2^{8}\ ,
\ee
which give $\ell_2=4$, $\ell_2^N=4$, $\ell_5=1$ and $\ell_5^N=3$. The degeneracy factors are
\be\label{F100}
\cF_1=1\ ,\qquad\qquad \cF_2=\cF_2'=1\ ,\qquad\qquad \cF_5=16\ ,\qquad\qquad\cF_5'=1\ .\vs{-5}
\ee
Note that $g$ acts cyclically on 3 copies of $5$-plet degenerate states in $(g^5,\uno)$ sector, i.e. $\mathfrak{b}=3$ in 
eq. \eqref{opintgm}, and leaves one state invariant $(\cF_5'=1)$. The coefficients $C_2$ and $C_5(1)$ are
\vs{-5}\be\label{C5C2hm100}
C_2=-4\ ,\qquad\qquad\qquad C_5(1)=-\sqrt5\ ,\vs{-5}
\ee
and $\theta_2=\theta_{5}^1=1$. All consistency conditions \eqref{opinttot} are satisfied. 

Next we consider the lattice $I=\tilde I(-1)$.
Since $P^2\in\bz$ for all $P\in I^*$, we can set $w=(0,0,0,0)$. In other words,  for the particular $I$ of this model,
all  generators of $\bz_2$ classes have integer norm.
Concerning the shift $v$, the modular invariance conditions \eqref{lmc_summ} require $10 v \in I$ and $5v^2 \in \bz$.
These are obviously satisfied by $v=0$, but to eliminate massless states in twisted sectors we have to choose a non-trivial
shift. Concretely, for absence of massless states in $g^\ell$-twisted sectors with $(\ell,10)=1$, it must be $v \notin I^*$.
With this choice, $2j v \notin  I^*$ as well, which guarantees no massless states in $g^{2j}$-twisted sectors. However, for 
the invariant lattice of this model there will always be massless vectors in the $g^5$-twisted sector. The reason is
that $P\cdot P' \in 2\bz$ for $P, P' \in I$, as seen from the Gram matrix in eq. \eqref{I100}. Thus, for any $v$
satisfying modular invariance, necessarily $5v \in I^*$. This means that $5v$ is always in a $\bz_2$ class of
$I^*$ and the momenta in the $g^5$-twisted sector are effectively not shifted.

Up to automorphisms of $I$ and translations in $I^*$ there is a unique choice of $v \notin I^*$, which can be taken
to be $v=\frac1{10}(0,1,3,0)$ in the $I$ basis. With this shift there are no massless states in all $g^k$-twisted sectors, except $k=5$. 
In the $g^5$-twisted sector, with $E_5=1$, there are massless states when $P_L=P_R=0$, $N_L=N_R=0$, 
$r^2=1$, and their multiplicity is $\frac2{10}(\cF_5+4\cF_5')=4$. All in all, the theory has 4 vector multiplets coming from 
twisted sectors, and its gauge group has a reduced rank 4.

It is important to note that one can further compactify the theory on an extra circle and add an appropriate shift vector along 
that circle to eliminate all massless states in twisted sectors. In this situation one obtains a theory of rank 1 (with one vector multiplet 
in the untwisted sector) in 5 dimensions.

\section{Final remarks}\label{section_final}

We have developed a systematic formalism to construct heterotic string vacua with 16 supercharges and reduced rank as 
$\bt^D/\bz_n$ asymmetric orbifolds. The key observation is that rank reduction requires absence of roots in
the normal sublattice under the $\bz_n$ automorphism of the $(16+D,D)$ momentum lattice $\Gamma$.
To fulfill this requirement we simply chose to assemble $\Gamma$ from sublattices of the Leech lattice.
Another possibility, considered e.g. in \cite{Chaudhuri:1995dj, Bossard:2017wum}, is to engineer $\Gamma$ from a Niemeier lattice with 
repeated factors and take the $\bz_n$ to be a cyclic 
permutation.\footnote{Starting with a Niemeier lattice, a more general procedure to obtain a normal sublattice without roots is 
to use an automorphism that does not belong to the Weyl group \cite[proposition 1.14.8]{Nikulin80int}.}

The derivation of the full orbifold partition function for order $n=2m$, with
$m$ an odd prime, was given in detail. Unlike the case of $n$ odd, when the order is even in general it is necessary 
to modify the standard orbifold action on $\Gamma$ in order to satisfy modular invariance.
Instead of doubling the order, as proposed e.g. in \cite{Narain:1990mw, Harvey:2017rko},
we were able to obtain a modular invariant partition function keeping an action of order $2m$.
This was achieved by introducing a vector characterizing the $\bz_2$ conjugacy classes of the dual invariant sublattice. 
Although we did not define the orbifold action globally on the full $\Gamma$, nor verified proper operator product expansions of 
vertex operators, we did check that untwisted sector states have positive integer multiplicities. 
We also proved that the partition function in all twisted sectors is consistent with the
operator interpretation, which in turn ensures positive integer multiplicities. To simplify expressions we specialized 
to invariant lattice of definite signature, but the extension to indefinite signature is straightforward and was considered
in appendix \ref{apprr}.

Besides absence of root vectors in the normal sublattice of $\Gamma$, rank reduction requires additional
conditions to avoid massless states in twisted sectors. We found that in some cases it was not possible to
satisfy these conditions once modular invariance was imposed. However, the problem can be evaded, i.e.
twisted states can be made massive, by adjoining a spectator circle and including an order $n$ translation along it 
in the orbifold action. Altogether, our formalism allows to construct fully supersymmetric heterotic vacua with diverse ranks in various
dimensions.  In particular, we presented $\bt^4/\bz_{2m}$ orbifolds, with rank $\frac12(5-m)^2$ for $m=1,3,5$, that had been 
conjectured in \cite{Fraiman:2022aik}. 
For the rank-0 case we actually encountered two distinct realizations, depending
on the choice of invariant sublattice, which do differ in the structure of massive states. 
Starting with a $\bt^6/\bz_{22}$ orbifold, we also discovered a heterotic archipelago in four dimensions, 
formed by islands without massless vector multiplets but with different massive matter content.

It would be interesting to further study the $6d$ heterotic islands, also in connection with the known $6d$ island 
\cite{Dabholkar:1998kv}, and others constructed as type II asymmetric orbifolds \cite{Fraiman:2022aik, Baykara:2025lhl}.
For the $4d$ archipelago it could be worthwhile to consider aspects of S-duality, as has been done in other CHL models 
\cite{Persson:2015jka, Persson:2017lkn}.

A natural follow-up is to examine dualities between the heterotic asymmetric orbifolds that we have constructed and 
compactifications of M-/F-theory. In some cases the strong coupling description is already known. For instance, the $\bt^4/\bz_7$ 
with rank 2, outlined at the end of subsection \ref{subsection_leech}, has a dual realization as M-theory on $\text{K3}\times S^1/\bz_7$. 
In fact, other rank-reduced heterotic orbifolds $\bt^4/\bz_{n}$, $n=2,\ldots,8$, are known to be dual to M-theory
on $\text{K3}\times S^1/\bz_n$, or type IIA on $\text{K3}$ with fluxes 
\cite{Schwarz:1995bj, Chaudhuri:1995dj, Chaudhuri:1995ee, deBoer:2001wca}. 
It is an open problem to find a concrete dual realization 
for the $\bt^4/\bz_2$ with rank 8 and the $\bt^4/\bz_6$ with rank 2. As indicated in \cite{Dabholkar:1998kv}, for the $6d$ and 
$4d$ islands the strong coupling limit is expected to be again an island, possibly differing in the massive spectrum.

Finally, our formalism may be applied to heterotic compactifications with less amount of supersymmetry by
considering automorphisms of $\Gamma$ that act on some or all right-moving directions. The corresponding
action on world-sheet fermions will break supersymmetry partially or completely. To achieve modular invariance
for automorphisms of even order, the orbifold action on $\Gamma$ might have to be modified as done
in this work, or in our previous study of non-supersymmetric $\bt^3/\bz_2$ orbifolds \cite{Acharya:2022shu}. 
We hope to report on this program in the near future.

\section*{Acknowledgments}
We thank C.~Angelantonj, K.~Budzik, X.~de~la~Ossa, B.~Fraiman, I.~Melnikov, M.~Montero, A.~Taormina, S.~Theisen and 
K.~Wendland for helpful conversations. 
We are specially grateful to B.~Acharya for collaboration at early stages of this work, and to H. Parra De Freitas for
many insightful comments, including remarks on the manuscript, as well as for pointing relevant references.
GA  thanks  CERN,  ICTP and IFT UAM-CSIC via the Centro de Excelencia Severo Ochoa for hospitality and support during completion of this paper. The work of G. A. is partially supported  by the PICT-2020-01760 grant.
AF thanks LAPth, CERN, ESI (within the programme ``The Landscape vs. the Swampland"), Bo\u{g}azi\c{c}i University, and ICTP, for their hospitality and support at various phases of this work.
IGZ would like to thank MITP for support and hospitality while parts of this work were done.

\appendix

\section{Existence of \texorpdfstring{$w$}{TEXT}}\label{app_w}

Consider the sublattice $I$ invariant under a $\bz_{2m}$ automorphism of the even self-dual lattice $\Gamma$. 
We have seen that $I^*$ has a number $2^{\ell_2}$ of $\bz_2$  and a number $m^{\ell_m}$ of  $\bz_m$ conjugacy 
classes -- cf. eq. \eqref{I*}. We will assume that $\ell_2$ is not zero. 
If $P \in \bz_2$ class, then $2P^2 \in\bz$ as $2P \in I$, therefore $P^2$ can be integer or half-integer.
We want to show that there exists a characteristic vector $w$ with $w \in I^*$ and $2w \in I$, such that
$P^2 =P\cdot w$ mod integer for all $P$ in a  $\bz_2$ class.
To this end, let us split the $\bz_2$ classes as $\bz_2^e$ and $\bz_2^o$, such that if $P \in \bz_2^e$, then $P^2$ is integer, and if 
$P \in \bz_2^o$, then $P^2$ is $\frac12$ mod integer. If there are no $\bz_2^o$ classes then $P^2\in\bz$ and $w$ can be chosen 
to be zero.  Then we will further assume henceforth that some $\bz_2^o$ class exists.

It is easy to show that there is an additive structure in the $\bz_2^e$ and $\bz_2^o$ classes, namely
\be
P_1, P_2 \in \bz_2^e \ \Rightarrow P_1 + P_2 \in \bz_2^e, \quad  P_1, P_2 \in \bz_2^o \ \Rightarrow \ P_1 + P_2 \in \bz_2^e, 
\quad P_1 \in \bz_2^e,  P_2 \in \bz_2^o \ \Rightarrow P_1 + P_2 \in \bz_2^o \, . 
\ee
This shows that the number of $\bz_2^o$ and $\bz_2^e$ classes will be equal. 
Let us denote by $P_o$ an element of some  $\bz_2^o$ class.  Now consider a sublattice $I^*_e$ of $I^*$ which consists of all 
the $\bz_m$ classes, as well as $\bz_2^e$ classes, and this implies that it also contains $I$. Clearly $I^*_e $ is a sublattice of $I^*$ 
and indeed, $I^*/I^*_e$ is $\bz_2$ and is explicitly given by $(0, P_o)$. Then $I$, which is $(I^*)^*$, will be a proper 
subset of $(I^*_e)^*$, because  $I^*_e$ is smaller than $I^*$. Furthermore, $(I^*_e)^*$ is a proper subset of $I^*$ 
because $I$ is contained in $I^*_e$. Thus, $I \subset (I^*_e)^* \subset I^*$, where both subsets are proper subsets. 
Now we can take $w$ to be an element of  $(I^*_e)^*$ such that it is not in $I$. Indeed, $P_o\cdot w$ cannot be integer, since 
if it was so, then $w \in I$. If it is not integer, then $P_o\cdot w$ must be $\frac12$ mod integer, 
since $2 P_o \in I$ and $w \in  (I^*_e)^* \subset I^*$.

In the case of $\bz_2$ automorphisms a similar construction of $w$ was presented in \cite{Acharya:2022shu}.
A simple instructive example is $I=A_1$, in which $w$ can be taken to be the fundamental weight of $SU(2)$.

\section{\texorpdfstring{$Z(g^2,g^{-1})$}{TEXT}}\label{app_g2}

In this appendix we compute the contribution of the right-moving compact directions to
$Z(g^2,g^{-1})$. We start from $Z(g,\uno)$ and obtain the $Z(g,g^2)$ sector 
by 2 $T$-transformations, as given in eq. \eqref{zggjb}. We denote this contribution to the partition function by
\vs{-5}\be\label{g2latt}
Z_{\rm lat}(g,g^2):=\frac1{\bar\eta^{2d}}\sum_{P\, \in \, I^*} \bar q^{\frac12 (P+v)_R^2} \op e^{2i\pi (P+v)^2} \ .
\ee
Recall that $I$ is purely right-moving and has dimension $2d$.
Next we write $P= X+Y$ for some $X \in \bz_2$ class and $Y\in \bz_m$ class of $I^*/I$. 
It is convenient to expand $Y$ in a basis of generators which can be taken to be the $f_{m,i}^I$ introduced in
the decomposition of $\Gamma$ in \eqref{Gamma}. To simplify notation we define $f_i:=f_{m,i}^I$ and write
\vs{-5}\be\label{P2}
Y= \sum_{i=1}^{\ell_m}  b_i f_i\ ,\vs{-5}
\ee
where  $\ell_m$ is the number of $\bz_m$ generators and $b_i$ for each $i$ runs from $0$ to $(m-1)$. 
The sum over $P$ becomes a double sum over $X$ and $Y$.
The sum over $Y$, which is actually over the $b_i$, is finite.
On the other hand, the sum over  $X$ will include all of $I$ and the $\bz_2$ generators of $I^*/I$, and is then an infinite sum. 
Next consider $P^2= X^2+Y^2+2 X\cdot Y$. 
Since $2 X \in I$, $2 X\cdot Y \in \bz$. Furthermore, $X^2=  X\cdot w$ mod integers -- see eq. \eqref{w} and appendix \ref{app_w}. 
Therefore, $P^2= X\cdot w+ Y^2$ mod integers.
Thus, the lattice part in $(g,g^2)$ can be expressed as
\vs{-5}\be\label{g2lattb}
Z_{\rm lat}(g,g^2)=\frac1{\bar\eta^{2d}} \sum_{Y} F(Y)\op e^{2 \pi iY^2}e^{2\pi iv^2}\ ,
\ee
where 
\be\label{fY}
F(Y):= \sum_{\substack{X \in Z_2 \\ {\rm classes\,of\,} I^*}}
\bar{q}^{\frac{1}{2}(X+Y+v)_R^2} \, e^{2 \pi i(2v+ w)\cdot (X+Y)}\ .
\ee
In the above we dropped $ e^{-2\pi i Y\cdot w}$ because $Y\cdot w \in\bz$ for all $Y$.
The reason is that $w$ is in a $\bz_2$ class while $Y$ is in a $\bz_m$ class of $I^*/I$, and $m$ is odd. 

We next perform a Poisson resummation on $F(Y)$. The dual lattice is the lattice of $\bz_m$ classes of $I^*$ and we obtain
\vs{-5}\be\label{fYpr}
F(Y) =  \frac{1}{\rm{vol}_2}\; \frac1{(i\bar\tau)^{2d}}
\!\!\!\!\! \sum_{\substack{P' \in Z_m \\ {\rm classes\,of\,} I^*}}
 \bar{q}^{\prime \frac{1}{2}(P'+2 v+ w)_R^2} e^{-2\pi i P'\cdot Y} e^{-2 \pi i(2v+ w+P')\cdot v}\ ,
\ee
where ${\bar q}^\prime=e^{2\pi i /\bar \tau}$. The volume $\rm{vol}_2$ of a basic cell of $\bz_2$ classes in $I^*$ will be 
evaluated shortly. Substituting eqs. \eqref{fYpr} in eq. \eqref{g2lattb}, and doing $\tau \to -1/\tau$, we find that in $Z(g^2,g^{-1})$ 
the lattice part is
\be\label{g2lattc}  
\!\!\!Z_{\rm lat}(g^2,g^{-1})=\frac{1}{\rm{vol}_2}\;  \frac1{\bar\eta^{2d}}\!\!\!\!\!
\sum_{\substack{P' \in Z_m \\ {\rm classes\,of\,} I^*}}\!\!\!\!\!\! \bar{q}^{\frac{1}{2}(P'+2 v+ w)_R^2} 
e^{-2 \pi i(P'+2v+ w)\cdot v} e^{2\pi i v^2}\!\!\!\!\!\!\!\!
\sum_{\substack{Y \in Z_m \\ {\rm classes\,of\,} I^*/I}}\!\!\!\!\!\!\!\! e^{-2\pi i P'\cdot Y} e^{2\pi iY^2}\ .
\end{equation}
Recall that sum over $Y$ means finite sums from $0$ to $(m-1)$ for each $b_i$ in \eqref{P2}. 

We now evaluate ${\rm{vol}}_2$. The volume of a cell in $I$ is $\sqrt{|I^*/I|}=\sqrt{2^{\ell_2}m^{\ell_m}}$ -- cf. eq. \eqref{I*}. 
Since in a given cell of $I$ there are  $2^{\ell_2} m^{\ell_m}$ basic cells of $I^*$, the volume of
a basic cell of $I^*$ will be $1/\sqrt{2^{\ell_2} m^{\ell_m}}$. The basic cell of $\bz_2$ classes in $I^*$ must  
include $m^{\ell_m}$ basic cells of $I^*$. Therefore
\vs{-5}\be\label{vol2}
{\rm{vol}}_2=\frac{m^{\ell_m}}{\sqrt{2^{\ell_2} m^{\ell_m}}}=\sqrt{\frac{m^{\ell_m}}{ 2^{\ell_2}}}\ . \vs{-5}
\ee

We can repeat the same steps and obtain the lattice part of the partition function for $(g^{2j}, g^{-1})$ sectors, where $1\le j\le m-1$. 
The result is
\be\label{g2alatt}  
\!\!\!Z_{\rm lat}(g^{2j},g^{-1})=\frac{1}{\rm{vol}_2}\;  \frac1{\bar\eta^{2d}}\!\!\!\!\!
\sum_{\substack{P' \in Z_m \\ {\rm classes\,of\,} I^*}}\!\!\!\!\!\! \bar{q}^{\frac{1}{2}(P'+2 jv+jw)_R^2} e^{-2 \pi i(P'+2jv+jw)\cdot v}
e^{2\pi i jv^2}\!\!\!\!\!\!\sum_{\substack{Y \in Z_m \\ {\rm classes\,of\,} I^*/I}} \!\!\!\!\!\!e^{-2\pi i P'\cdot Y}  e^{2\pi ijY^2}\ .\vs{-5}
\end{equation}
This expression may be simplified further to remove the double sum as we now explain.

Consider the sum $\sum_{Y} e^{-2\pi i P'\cdot Y} e^{2 \pi i j Y^2}$ appearing above. 
We will try to determine its $P'$-dependence. 
Since $P'$ is in a $\bz_m$ classes of $I^*$, to begin it can be expanded as  $\sum_i  b_i f_i$ mod $I$, where the $f_i$
are the same generators used in eq. \eqref{P2} and each $b_i$ takes values in  $\{0,1,\ldots,m-1\}$. 
But then we can equally well choose a different basis where $f_i$ are replaced 
by any even number times $f_i$, so long as this even number is coprime with $m$. In particular, we can choose the basis where 
$f_i$ is replaced by $2 j f_i$, for $j=\{1,2,\ldots,m-1\}$. 
What this means is that $P'$ can be expressed as $2 j\sum_i   b_i f_i$ mod $I$. 
For conciseness we define
\vs{-5}\be\label{bifib}
b_i f_i := \sum_{i=1}^{\ell_m} b_i f_i \ ,\qquad\qquad b_i\in\{0,\ldots,m-1\}\ .\vs{-5}
\ee
Then
\begin{eqnarray}
  \sum_{Y} e^{-2\pi i P'\cdot Y} e^{2 \pi i j Y^2} &&\hspace{-18pt}=  \sum_{Y} e^{-2\pi i (2j\, b_i f_i\cdot Y)} e^{i \pi 2j Y^2}
                                                  = e^{-2\pi i j(b_i f_i)^2} \sum_{Y} e^{2j \pi i  (Y-b_i f_i)^2}\nonumber\\
                                                  &&\hspace{-18pt}= e^{-2\pi i j(b_i f_i)^2} \sum_{Y} e^{2 \pi i j Y^2} = C_m(j) e^{-2\pi i j (b_i f_i)^2}\ ,
\end{eqnarray}
where in the last line we shifted $Y \rightarrow Y+b_i f_i$, which is just a redefinition of the coefficients
in \eqref{P2}, and
\be\label{Cma}
C_m(j) :=\hspace{-15pt}\sum_{\substack{Y \in Z_m \\ {\rm classes\,of\,} I^*/I}}\hspace{-15pt} e^{2 \pi i j Y^2}\ .
\ee
Thus eq. \eqref{g2alatt} becomes
\begin{equation}\label{g2alattb}
 Z_{\rm lat}(g^{2j},g^{-1})= \frac{1}{\rm{vol}_2}\frac{C_m(j)\op e^{2\pi i j  v^2}}{\bar\eta^{2d}}\sum_{b_i=0}^{m-1}
\sum_{P \in I} \bar{q}^{\frac{1}{2}(P+2 j b_i f_i+2j v+ j w)_R^2} e^{-2 \pi i(P+2j b_i f_i+2j v+ j w)\cdot v} e^{-2\pi i j(b_i f_i)^2}\ .
\end{equation}
We next compute $C_m(j)$.

\subsection{\texorpdfstring{$C_m(j)$}{TEXT}}\label{appCma}

We can compute $C_m(j)$ up to a phase. Consider $|C_m(j)|^2$ (in the following $P$ and $P'$ are in $\bz_m$ classes of $I^*/I$) 
\begin{eqnarray}\label{Cma2}
  |C_m(j)|^2 =\!  \sum_{P'} \sum_{P} e^{2 \pi i j P^2}e^{-2 \pi i j P'^2} \!=\!  \sum_{P'} \sum_{P} e^{2 \pi i j (P+P')(P-P')} 
          \!=\!  \sum_{P}  e^{2 \pi i j P^2}\sum_{P'}e^{-4\pi i j P\cdot P'}\ ,\;\;
\end{eqnarray}
where in the last equality we have redefined $P \rightarrow P-P'$. Now let us look at the term $\sum_{P'}e^{-4\pi i j P\cdot P'}$ above. 
Suppose $P$ is a non-trivial element in $\bz_m$ classes (i.e. it is not in $I$). If $P \notin I$, then there must be some element in, 
say $P_1 \in\bz_m$ class, such that $P\cdot P_1$ is not an integer (otherwise $P$ will be trivial, i.e. in $I$, since $P$ being in a $\bz_m$ 
class will have integer dot products with all the $\bz_2$ classes also, which would imply that $P$ dotted with any element of $I^*$ is 
integer). 
As before we now choose a basis $f_i$, $i=1, \ldots, \ell_m$, for the $\bz_m$ generators and
expand $P'=\sum_{i=1}^{\ell_m} b_i f_i$, where $b_i\in\{0,\ldots,m-1\}$. 
Then, if $P \notin I$
\begin{equation}\label{PPp}
  \sum_{P'}e^{-4\pi i j  P.P'} =  \prod_{i=1}^{\ell_m} \sum_{b_i=0}^{m-1} e^{-4\pi i j b_i P\cdot f_i}\ .
  \end{equation}
Since $P\cdot f_1$ is not an integer, it must be  $k/m$ for some  nonzero integer $k$ coprime with $m$. We also know that $j$ is  
coprime with $m$, since $1\le j\le (m-1)$ and $m$ is odd prime. Then 
$\sum_{b_1=0}^{m-1} e^{4\pi i j b_1 \frac km} =0$. Hence, in eq. \eqref{Cma2}, $P$ must be trivial and hence the sum 
over $P$ just collapses to just one term namely $P=0$ , but then the  sum over $P'$ gives $m^{\ell_m}$. 
We then conclude that $C_m(j) = \sqrt{m^{\ell_m}}$ up to a phase, i.e. 
\vs{-5}\be\label{Cmab}
C_m(j):=e^{i\pi\theta_m^j}\,\sqrt{m^{\ell_m}}\ .
\ee
A quantity that will appear in partition functions is
\be\label{phaseCm}
\frac{C_m(j)}{\text{vol}_2} = e^{i\pi \theta_m^j} \sqrt{2^{\ell_2}}\ ,\vs{-5}
\ee
where we used eq. \eqref{vol2}.

We next define two auxiliary sums 
which will be useful in our computations. Consider an invariant lattice of signature $(d_L, d_R)$ and $I^*/I=\bz_2^{\ell_2} \times \bz_m^{\ell_m}$. Milgram's theorem gives \cite[appendix 4]{Milnor73}
\be\label{mil1}
C \ = \sum_{P \in I^*\!/I} e^{i \pi P^2} = \sqrt{2^{\ell_2}\,m^{\ell_m}}e^{\frac{i\pi}{4}(d_L - d_R)} \ .
\ee
We next write $P=X+Y$, with $X \in \bz_2$ class and $Y \in \bz_m$ class. Using $X \cdot Y \in \bz$, we find
\be
C \ =  \hspace*{-4mm} \sum_{\substack{X\, \in \, \bz_2\\ \text{classes of}\, I^*/I}}\hspace*{-6mm}  e^{i \pi X^2}  \!\!
\sum_{\substack{Y\, \in \, \bz_m\\ \text{classes of}\, I^*/I}} \hspace*{-6mm}  e^{i \pi Y^2}  = B_2\; B_m \ ,\vs{-5}
\ee
where we have defined
\vs{-5}\be\label{B2Bma}
B_2 \ := \hspace*{-4mm} \sum_{\substack{X\, \in \, \bz_2\\ \text{classes of}\, I^*/I}} \hspace*{-6mm}  e^{i \pi X^2}  \ , \qquad\qquad\qquad
B_m\ :=  \hspace*{-4mm} \sum_{\substack{Y\, \in \, \bz_m\\ \text{classes of}\, I^*/I}}\hspace*{-6mm}   e^{i \pi Y^2} \ .
\ee
Following the same arguments we had above for $C_m(j)$, we can prove that $|B_2|^2 = 2^{\ell_2}$ and $|B_m|^2 = m^{\ell_m}$. 
We can then write
\be\label{B2Bm}
B_2 := e^{i\pi \phi_2} \sqrt{2^{\ell_2}} \ , \qquad\quad  B_m := e^{i\pi \phi_m} \sqrt{m^{\ell_m}} \ .
\ee
According to Milgram's theorem, cf. \eqref{mil1}, the phases in $B_2$ and $B_m$ then satisfy
\be\label{milgamma}
e^{i\pi \phi_2} = e^{\frac{i\pi}{4}(d_L - d_R)}  e^{-i\pi \phi_m} \ .\vs{-5}
\ee
This relation will be helpful later.

The auxiliary quantity $B_m$ turns out to be related to the $C_m(1)$ entering in $Z(g^2,g^{-1})$
according to $C_m(1)^2=B_m^2$. 
To prove this we substitute the expansion of $Y$ in \eqref{P2} into eqs. \eqref{Cma} and \eqref{B2Bma} to obtain
\be
  C_m(1)= \sum_{b_1=0}^{m-1}\cdots\sum_{b_{\ell_m}=0}^{m-1} e^{2\pi i (b_i f_i)^2}\ , \qquad\qquad
  B_m= \sum_{b_1=0}^{m-1}\cdots\sum_{b_{\ell_m}=0}^{m-1} e^{i\pi (b_i f_i)^2} \, .
         \ee
with the convention in \eqref{bifib}. We will also use the short hand notation $\sum_{b_i}\!\!:=\sum_{b_1=0}^{m-1}\cdots\sum_{b_{\ell_m}=0}^{m-1}$. Then
\vs{-5}\be
C_m(1)^2 =\sum_{b_i}  \sum_{b'_i}  e^{2\pi i (b_i f_i)^2+2\pi i (b'_i f_i)^2} \!
=\sum_{b_i}  \sum_{b'_i} e^{i\pi \big(( b_i+b'_i) f_i\big)^2} e^{i\pi  \big(( b_i-b'_i) f_i\big)^2}\ .\vs{-5}
  \ee
Now call $ l_i:=b_i-b'_i$. Then $b_i+b'_i= l_i+ 2 b'_i =: l'_i$. Clearly as $b_i$ and $b'_i$ vary between $0$ and $(m-1)$, $l_i$ will span all integers $0$ to $(m-1)$ mod $m$. Then for a fixed $l_i$, $l'_i$ will range over all integers mod $m$ that are $l_i$ mod even numbers $2 b'_i$. But for $m$ odd, which is the case we are considering, $2b'_i$ will span all integers modulo $m$. 
Thus, for odd $m$, the original sums over $b_i$ and $b'_i$ from $0$ to $(m-1)$ reduce to independent sums over $l_i$ and $l'_i$ from 
$0$ to $(m-1)$ and in the end
\vs{-5}\begin{equation}\label{Cm1Bm}
C_m(1)^2 =\sum_{l_i}  \sum_{l'_i}  e^{i\pi ( l_i f_i)^2}e^{i\pi (l'_i f_i)^2}= B_m^2\ .
\end{equation}
Likewise it can be shown that $C_m(2l)^2 = C_m(1)^2$.
It then follows that
\be\label{them1phi}
2\theta_m^1=2\theta_m^{2l}\!\!\!\!\mod2=2\phi_m\!\!\!\!\mod2\ ,
\ee
where we used  the definitions of $C_m(j)$ and $B_m$ in eqs. \eqref{Cmab} and \eqref{B2Bm}. 

From the definition of $C_m(j)$ in eq. \eqref{Cma}, and the fact that $mY^2 \in 2\bz$ for all $Y$ in $\bz_m$ classes
of $I^*/I$, we deduce that $C_m(m-1)= C_m^*(1)$. But since $(m-1)$ is even this then implies $C_m(1)^2={C^*_m(1)}^2$,
and in turn 
\be\label{thetams}
4\theta_m^1 \in 2\bz, \qquad\qquad 4\theta_m^{2l} \in 2\bz, \qquad\qquad 4\phi_m \in 2\bz \, .
\ee
Similarly, $C_m(m-2k)= C_m^*(2k)$ yields $\theta_m^{m-2k} = - \theta_m^{2k}\, \text{mod} \, 2$. 
Using that $(m-2k)$ is odd we can then show that  $2\theta_m^{2l+1}=2\theta_m^1 \, \text{mod}\, 2$.

\section{\texorpdfstring{$Z(g^m,g^{-(2j+1)})$, $2j+1\ne m$}{TEXT}}\label{app_gm}

Consider $Z(g^{\ell},1)$, for $\ell$ an odd number not equal to $m$, as computed in eq. \eqref{zgl1c}. 
Performing a $T^m$-transformation leads to $Z(g^{\ell},g^m)$ in which the contribution of the right-moving compact directions is
\be\label{gmlatt}
Z_{\rm lat}(g^\ell,g^m):=\frac1{\bar\eta^{2d}}\sum_{P\, \in \, I^*} \bar q^{\frac12 (P+\ell v)_R^2} \op e^{i\pi m (P+\ell v)^2} \ .
\ee
For $P$ in $I^*$ we again write $P=X+Y$, 
but now we expand $X$ in a way similar to eq. \eqref{P2}, namely $X=\sum_i a_i f^I_{2,i}$, 
where  $f^I_{2,i}$ are the $\bz_2$ generators of $I^*/I$ introduced in the decomposition of $\Gamma$, $i=\{1,\cdots, \ell_2\}$, 
and $a_i$ runs from values $0$ to $1$. Thus the sum over $X$ will be finite.
On the other hand, the sum over $Y$ will include all the vectors of $I$ and $\bz_m$ classes of $I^*/I$, and will be infinite.
In the phase $e^{i\pi m (P+\ell v)^2}$ the momentum dependent exponent involves $m(X^2+Y^2+2X\cdot Y)$, 
which reduces to $m X^2$ mod 2, because $m Y \in I$ and therefore $X\cdot m Y$ is integer, and $m Y^2$ is even. 
Doing the Poisson resummation as before, the lattice part of the $(g^m,g^{-\ell})$ sector is found to be
\begin{equation}\label{km0}  
 Z_{\rm lat}{(g^m,g^{-\ell})}= \frac{1}{\rm{vol}_m}\,\frac{1}{\bar{\eta}^{2d}}\!\!\!\!\!
 \sum_{\substack{P' \in Z_2 \\ {\rm classes\,of\,} I^*}}\!\!\!\!\!
\bar{q}^{\frac{1}{2}(P'+m \ell v)_R^2} e^{-2 \pi i(P'+m \ell v)\cdot\ell v}\!\!\!\!\!\!\!
\sum_{\substack{X \in Z_2 \\ {\rm classes\,of\,} I^*/I}}\!\!\!\!\!\!\! e^{-2\pi i P'\cdot X}  e^{i \pi m X^2}e^{i\pi m \ell^2 v^2}\ .
\end{equation}
The volume of the basic cell of $\bz_m$ classes in $I^*$ can be evaluated as explained earlier for ${\rm{vol}_2}$.
We find
\begin{equation}\label{volm}
\text{vol}_m= \sqrt{\frac{2^{\ell_2}}{m^{\ell_m}}} \ .
\end{equation}

Since $\ell$ is odd, i.e. $\ell=2j+1$ (excluding $\ell=m$) and $2m v \in \bz_2$ class of $I^*$, and since $P' \in \bz_2$ class of $I^*$, 
we can absorb $2m v$ in $P'$ by shifting $P' \rightarrow P'-2 jm v$ to recast \eqref{km0} as
\begin{equation}\label{km}  
Z_{\rm lat}{(g^m,g^{-(2j+1)})}\!=\! \frac{1}{\rm{vol}_m} \frac{1}{\bar{\eta}^{2d}}\!\!\!\!\!\!
\sum_{\substack{P' \in Z_2 \\ {\rm classes\,of\,} I^*}}\!\!\!\!\!\!
\bar{q}^{\frac{1}{2}(P'+m  v)_R^2} e^{-2 \pi i(P'+m  v)\cdot\ell v}  e^{\pi i m (2l+1)^2 v^2} \!\!\!\!\!\!\!\!\!
\sum_{\substack{X \in Z_2 \\ {\rm classes\,of\,} I^*/I}}\!\!\!\!\!\!\!\!\! 
e^{-2\pi i( P' - jw)\cdot X}  e^{i \pi m X^2} \ .
\end{equation}
Here we used that $e^{2\pi i j 2m v\cdot X}= e^{2\pi ij w\cdot X}$, since $2mv=w$ mod $I$.

Now we want to express the above double sum involving $P'$ and $X$ as disentangled sums. Since $m$ is odd, we can always 
express it as $m=\epsilon \, \text{mod} \, 4$, where $\epsilon= \pm 1$, and so $\epsilon$ is determined by $m$. 
Then $e^{i \pi m X^2}=   e^{i  \pi\epsilon X^2}$, since $2 X^2$ is integer because $X$ is in a $\bz_2$ class. 
In the sum over $X$ we can then complete squares in the exponent and shift $X \rightarrow X+\epsilon(P'-jw)$
since $\epsilon(P'-jw)$ is also  in $\bz_2$ classes. In the end the two sums in eq. \eqref{km} decouple and
the final result is
\begin{equation}\label{kmf}  
Z_{\rm lat}{(g^m,g^{-(2j+1)})}= \frac{C_2}{\rm{vol}_m}\,\frac{e^{i\pi m(2j+1)^2v^2}}{\bar\eta^{2d}}\!\!\!\!\!\!
\sum_{\substack{P' \in Z_2 \\ {\rm classes\,of\,} I^*}}\!\!\!\!\! \bar{q}^{\frac{1}{2}(P'+m  v)_R^2} e^{-2 \pi i(P'+m  v)\cdot (2j+1) v}e^{-i\pi \epsilon (P'-jw)^2}\ ,
\end{equation}
where
\begin{equation}\label{C2}
C_2:=\!\!\!\!\!\!\!\! \sum_{\substack{X \in Z_2 \\ {\rm classes\,of\,} I^*/I}}\!\!\!\!\!\!\! e^{i\pi \epsilon X^2} \, .
\end{equation}
We can repeat the arguments in appendix \ref{appCma} and prove that $C_2= \sqrt{2^{\ell_2}}$ up to a phase, i.e.
\be
\label{c2volmvalue}
C_2:= e^{i\pi \theta_2} \sqrt{2^{\ell_2}}\ .
\ee
Below we will show that $\theta_2$ can be related to $w^2$.

\subsection{\texorpdfstring{$\theta_2$}{TEXT}}\label{appC2}

Recall the definition \eqref{w}, i.e.
\vs{-5}\be\label{winexp}
e^{2i\pi X^2} = e^{2 \pi i X\cdot w}\;\;\forall X \in \bz_2~{\rm classes~of}~I^*\ ,\vs{-5}
\ee
where $w$ itself is in a $\bz_2$ class of $I^*$. Thus, we may write $C_2$ in eq. \eqref{C2} as 
\be
C_2 =\!\!\! \hspace*{-4mm} \sum_{\substack{X\, \in \, \bz_2\\ \text{classes~of}\, I^*/I}} \hspace*{-6mm}  e^{i \pi\epsilon (X-w)^2}  =
e^{i\pi\epsilon w^2} C^*_2 \ ,
\ee
where we used \eqref{winexp} in the second equality.
But $C_2 = e^{i\pi \theta_2} \sqrt{2^{\ell_2}}$. Therefore $e^{i\pi\epsilon w^2} = e^{2i\pi \theta_2}$. A similar computation 
for $B_2$ in \eqref{B2Bm} yields $e^{i\pi w^2} = e^{2i\pi \phi_2}$. Thus
\vs{-5}\be\label{w2th2}
w^2=2\epsilon\,\theta_2\!\!\!\mod2\ , \qquad\qquad w^2=2\,\phi_2\!\!\!\mod2\ .
\ee 
Now, notice that eq. \eqref{milgamma} requires $\phi_2+\phi_m+\frac{d_R-d_L}4\in2\bz$. 
This then implies
\vs{-5}\be\label{milgammab}
w^2+2\theta_m^1+\frac{d_R-d_L}2\in2\bz\ ,
\ee
after inserting the identities \eqref{them1phi} and \eqref{w2th2}. In the case discussed in the main text the invariant lattice
has signature $(0,2d)$, i.e. $d_L=0$ and $d_R=2d$. Since $2\theta_m^1 \in \bz$, cf. \eqref{thetams}, 
we then conclude that $w^2 \in \bz$ is a property of $I$ of signature $(0,2d)$.

\section{Another form for \texorpdfstring{$Z_{\rm lat}(g^j,g^{-1})$}{TEXT}}\label{app_Lsum}

In this appendix we want to consider again the contribution of the right-moving compact directions to $Z(g,g^j)$ in eq. \eqref{zggjb}, now for generic $j$. It reads 
\vs{-5}\be\label{gjlatt}
Z_{\rm lat}(g,g^j) :=\frac1{\bar\eta^{2d}}\sum_{P\, \in \, I^*} \bar q^{\frac12 (P+v)_R^2} \op e^{i\pi j (P+v)^2} \ .
\ee
To derive $Z_{\rm lat}(g^j,g^{-1})$ we need to do a $\tau \to -1/\tau$ transformation but Poisson resummation of the
lattice sum cannot be done directly because the exponent in the phase involves $P^2$. In the preceding appendices we
proceeded by decomposing $P \in I^*$ in a way suited to the case where $I^*$ only has $\bz_2$ and $\bz_m$ 
conjugacy classes. More generally we can just make the change of variables $P=P' + Q$, then sum $P'$ over $I$ and
$Q$ over $I^*/I$ so that the original $P$ is summed over all of $I^*$. The simplification arises because $P^2=Q^2 \, \text{mod} \, 2$,
and then we can do Poisson resummation over $P'$ in the standard way (see e.g. appendix 9.6 in  \cite{Blumenhagen:2013fgp}).
In this manner we arrive at
\vs{-5}\be\label{trick1}
Z_{\rm lat}(g^j, g^{-1})=\frac{e^{i\pi j v^2}}{\bar\eta^{2d}}\sum_{P\, \in \, I^*} 
\bar q^{\frac12 (P+jv)_R^2} \op e^{-2i\pi (P+jv)\cdot v} \op L(j, P) \ ,
\ee
where
\vs{-5}\be\label{lsumj}
L(j,P)=\frac{1}{\sqrt{|I^*/I|}}\sum_{Q\, \in \, I^*/I} 
e^{i\pi j Q^2} \op e^{-2 i \pi P\cdot Q}  \ .
\ee
The sum over $I^*$ in eq. \eqref{trick1} can again be traded by an infinite sum over $I$ and a finite sum over $I^*/I$.
In the end we obtain
\vs{-5}\be\label{trick2}
Z_{\rm lat}(g^j, g^{-1})=\frac{e^{i\pi j v^2}}{\bar\eta^{2d}} \sum_{\tilde Q\, \in \, I^*/I} L(j,\widetilde Q)
\sum_{P\, \in \, I} \bar q^{\frac12 (P+\widetilde Q + jv)_R^2} \op e^{-2i\pi (P+\widetilde Q + jv)\cdot v}\ .
\ee
Notice that the sum in $L(j,\widetilde Q)$ is finite and for a given $I$ it can be evaluated explicitly.

The disadvantage of the final expression in eq. \eqref{trick2} is that the spectrum is not immediately visible.
It is revealed only after doing the finite sum in $L(j,\widetilde Q)$. For instance, when $I^*$ only has $\bz_2$ and $\bz_m$
classes, we find that $L(j,\widetilde Q)$ for $j=m$ is different from zero only if $\widetilde Q$ belongs to $\bz_2$ classes of $I^*/I$, in 
agreement with our previous result in eq. \eqref{km}. For $j=2$ we instead find that $L(j,\widetilde Q + w)$ is different
from zero only if $\widetilde Q$ belongs to $\bz_m$ classes of $I^*/I$, as obtained before, cf.  eq. \eqref{g2alattb}.
For particular examples the full expressions can be matched in detail.

\section{Non-equivalent shift vectors}\label{appnoneqv}

In this appendix we comment about the requirement \eqref{vequiv} for two shift vectors to be non-equivalent and thus lead to 
different models. For simplicity let us denote $\theta v_2=v'_2$.

Notice first that for $z=1$,  if $v_1-v'_2=Q \in I^*$,  then the untwisted sectors will be the same since the only difference 
resides in the projectors, cf. eq. \eqref{z1g}. Namely, in the $(\uno,g^\ell)$ subsector we will have 
$e^{2i\pi P\cdot\ell v_1}=e^{2i\pi P\cdot\ell (v'_2+Q)}=e^{2i\pi P\cdot\ell v'_2},$ since  $P\in I$ and therefore $P\cdot Q\in\mathbb{Z}$.
However, it could happen that even if eq. \eqref{vequiv} is satisfied for $z=1$, we can still have $z\ne 1$ such that $zv_1-v'_2=Q \in I^*$.
The full effect is a reordering of the different terms in the sum over $\ell=0,\dots 2m-1$. 
Namely, $\ell v'_2\rightarrow ({\ell z\,{\rm mod}\ 2m}) v_1$  and therefore the sum remains invariant. 
Notice that for $\ell=m$,  $mv'_2= mzv_1-mQ=m(z-1)v_1+mv_1-mQ=mv_1+Q'$ with $Q'\in I^*$, since $2mv\in I^*$ and 
therefore the sector $(\uno,g^m)$ is the same for both shifts.

A similar situation occurs for twisted sectors.
In a $(g^\ell,g^k)$ subsector the shift dependent term is of the form 
${\bar q}^{\frac12(P+\ell v+w\delta_{\ell,2\,{\rm mod}\, 4})^2}{e}^{2i\pi\frac12(P+\ell v+w\delta_{\ell ,2\,{\rm mod} 4})^2}$.
For $z=1$ and $v_1-v'_2=Q \in I^*$ the difference can be absorbed in a redefinition of $P$ in the corresponding dual lattice. 
In the second situation the $g^\ell$ and $g^{\ell z\,{\rm mod}\ 2m}$ sectors will be exchanged but the sum over $\ell$ will be invariant.
In particular it can be seen that an $\ell$ even sector is exchanged with an $\ell \, {\rm mod}\, 4$ sector and that the  
$g^m$ sector remains invariant.

\section{Examples with indefinite invariant lattices}\label{apprr}

In this appendix we consider $\bt^4/\bz_{2m}$ asymmetric orbifolds in which the automorphism leaves
invariant some left-moving, as well as all the right-moving, directions of $\Gamma$. The analysis in sections
\ref{section_Z} and \ref{section_op} can be generalized making necessary changes that do no alter the main points.
We present two models which correspond to entries HM5 and HM63 of \cite[Table 1]{HM}, with respectively $\bz_2$ and 
$\bz_6$ automorphisms. In both cases $\Gamma$ has signature $(20,4)$, the normal lattice is a sublattice of the Leech lattice with 
signature $(s,0)$, and the invariant lattice has signature $(20-s,4)$, where $s=12$ and $s=18$ for the HM5 and HM63,
respectively.

Since now the invariant lattice has left-moving directions, there are massless states in vector multiplets in the untwisted sector, and so 
these orbifolds are not islands. However, we show that there are no massless states in the twisted sectors and hence, the rank of the 
gauge group is reduced from $20$ to $20-s$. The existence of $\bt^4/\bz_2$ and  $\bt^4/\bz_6$ heterotic orbifolds with
gauge group of reduced rank 8 and 2 respectively, was conjectured in \cite{Fraiman:2022aik}, where they were associated to the 
genus D and genus J of the classification in \cite{Hohn:2017dsm}. To our knowledge, the construction of these orbifolds is not known in the literature. In the following we will show how they can be built in the formalism developed in this paper.

\subsection{HM5}\label{apprr_hm5}

This is a $\bz_2$ asymmetric orbifold in 6 dimensions. The normal lattice $N$ has rank $s=12$ and the automorphism $\Theta$
in the basis of $N$ is just minus the identity matrix.
The eigenvalue distribution of $\Theta$ is then $t=\tfrac1{2}([1]^{6})$, corresponding to frame shape $2^{12}$ \cite{HL}.
The zero point energy is $E_1=\frac s{16}=\frac34$. 

In the HM5, the normal and invariant lattices of the Leech lattice happen to be isometric, specifically
$N\cong \tilde I\cong D_{12}^+(2)$, where $D_{12}^+$ is the odd self-dual lattice of rank 12 obtained adjoining the spinor class
to $D_{12}$. 
The invariant lattice $I$, which is correlated with $N$ to construct the even self-dual lattice $\Gamma(20,4)$, must have signature $(8,4)$ and
\be\label{hm5_disc}
I^*/I\cong N^*/N\cong \bz_2^{12}\ .\vs{-5}
\ee
Hence, $\ell_2\!=\!s=\!12$, and the degeneracy factor is $\cF_1=1$. Note that since $N^*$ has vectors with half-integer length 
squared, $I^*$ must also have $\bz_2$ classes with half-integer length squared to correlate with $N$ to form $\Gamma(20,4)$. 
This means that in this HM5 there will be a characteristic vector $w$ with the crucial property in eq. \eqref{w}.

It is known that indefinite even 2-elementary lattices of same signature, same discriminant group, and such that $P^2$ is not an integer 
for all $P \in I^*$, are all isometric \cite{Nikulin80int}. In our case this means that for the $I$ of signature $(8,4)$ there are infinite 
choices related by $O(8,4)$ transformations. Indeed, a generic $I$ will have 32 moduli that take continuous values.
For example, we find that the lattice $I_0=7A_1 \oplus U(2) \oplus 3 A_1(-1)$, with $I^*/I \cong \bz_2^{12}$, can be correlated with $N$.
This $I_0$ corresponds to a particular point of the moduli space of the generic $I$.
It is easy to guess other possibilities at different points of moduli space, e.g. $I_1=8A_1 \oplus 4A_1(-1)$, which was
proposed in \cite[Table 1]{Persson:2015jka}. In this example we can actually do better because 
in \cite[eq. (5.34)]{Acharya:2022shu} it was explicitly shown that $7A_1 \oplus U(2)$ is in the moduli space of a $\Upsilon(8,1)$
lattice whose moduli are a radius and a 7-dimensional Wilson line. Moreover, it was proven that at special values of these moduli
the $(8,1)$ lattice becomes $\Upsilon_1=8A_1 \oplus A_1(-1)$ or $\Upsilon_2=E_8(2) \oplus  A_1(-1)$.
The above mentioned $I_1$ is thus recovered and we also learn that $I_2=E_8(2) \oplus  4A_1(-1)$ is yet another $(8,4)$
lattice that can be correlated with $N \cong D_{12}^+(2)$. Below we will consider these two specific choices of $I$.

\subsubsection{Partition function}\label{pf_hm5}

The partition function can be readily obtained from the general expressions in section \ref{section_Z}.
We can equally adapt the detailed results of \cite{Acharya:2022shu} for asymmetric  $\bz_2$ orbifolds.
In either way we find that in the untwisted sector the lattice pieces simply read
\be\label{Zunt}
  Z(\uno,\uno)=\frac{1}{\eta^{20}\bar{\eta}^{4}} 
              \sum_{P\in \Gamma} q^{\frac{1}{2}P_L^2}\bar{q}^{\frac{1}{2}P_R^2}\ ,
\qquad         
 Z(\uno,g)=\frac{1}{\eta^{20}\bar{\eta}^{4}}   \left(\frac{2\eta^3}{\vartheta_2}\right)^{\!\! \frac s2}
 \sum_{P\in I}  q^{\frac{1}{2}P_L^2}
  \bar{q}^{\frac{1}{2}P_R^2} \, e^{2\pi i P\cdot v} \ . \vs{-5}
\ee
Modular transformations then give the twisted sector terms
\begin{align}\label{Ztw}
  Z(g,1)&=\frac{1}{\eta^{20}\bar{\eta}^{4}}
   \left(\frac{\eta^3}{\vartheta_4}\right)^{\!\! \frac s2}
              \sum_{P\in I^*} q^{\frac{1}{2}(P+v)_L^2}\bar{q}^{\frac{1}{2}(P+v)_R^2}\ ,\\
  Z(g,g)&=\frac{1}{\eta^{20}\bar{\eta}^{4}}
   \left(\frac{\eta^3}{\vartheta_3}\right)^{\!\! \frac s2}
    e^{i\pi(v^2 + \frac{s}8)}   \
\sum_{P\in I^*} q^{\frac{1}{2}(P+v)_L^2}\bar{q}^{\frac{1}{2}(P+v)_R^2} \,  
e^{2\pi i P\cdot v}  e^{i \pi P^2}  \ ,\nn 
\end{align}
where $s=12$. 
The contribution due to uncompactified light-cone world-sheet bosons and right-moving world-sheet fermions 
is the same in both untwisted and twisted sector, cf. eq. \eqref{zxzpsidef}.

Applying $\tau \to \tau + 1$ to $Z(g,g)$ will bring in a phase $e^{2 \pi i P^2}$ inside the sum that could spoil 
modular invariance. However, again we can show that there exists a characteristic vector $w$, with $w \in I^*$, $2 w\in I$,
such that $e^{2 \pi i P^2} = e^{2 \pi i P \cdot w}$ for all $P \in I^*$.
Modular invariance $Z(g,g^2) \!= \!Z(g,\uno)$ can then be fulfilled if
\be\label{condz2}
2v + w \in I\ , \qquad\qquad\qquad 2v^2 + \frac{s}4 \in 2\bz \ ,
\ee
which, as expected, agrees with the conditions in eq. \eqref{lmc_summ} upon setting $m=1$ and $E_1=s/16$.
Note that necessarily $v$ must be different from zero and moreover $v \notin I^*$.  In the examples below we will
determine explicit solutions for $v$.

From the full partition function we can obtain the spectrum of massless and massive states in the emerging
$6d$ theory with 16 supercharges. To this end we remark that eqs. \eqref{Zunt} and \eqref{Ztw} entail integer
multiplicity of states. For the untwisted sector the same analysis of subsection \ref{subsection_op_un} applies. 
In the twisted sector all level-matched states survive the orbifold projection with multiplicity one.

Generic features of the spectrum were already discussed in section \ref{section_hm}.
Let us concentrate on massless states and discuss first the simpler twisted sector. From the $Z(g,g^k)$ terms
we see that there are no massless states since $v \notin I^*$. 
In the untwisted sector the massless matter includes the gravity multiplet and a number of vector multiplets.
At generic points of moduli space of $I$ there are no momenta with $P_L^2=2$, massless vectors only
arise from the 8 oscillator modes along the left-moving invariant directions, and the gauge group will be
$U(1)^8$. At special points of moduli space there can be enhancement to a non-Abelian group of rank 8, as
exemplified below.

\subsubsection{Example 1}\label{ex1_hm5}

In this example we take invariant lattice $I_1 \!=\! 8 A_1 \oplus 4 A_1(-1)$. 
The characteristic vector $w$ can obtained as explained in appendix \ref{app_w}, see also \cite{Acharya:2022shu}.
In the $I_1$ basis it is given by
\be\label{whm51}
w = \frac1{2}(1,1,1,1,1,1,1,1;1,1,1,1)\ , \vs{-5}
\ee
where the semicolon separates the left- and right-moving components.
Note that $w^2$ is even. In fact, Milgram's theorem \eqref{mil1} leads to $w^2+\frac s2\in2\bz$ for an invariant lattice of 
signature $(20-s,4)$.
The unique choice for the shift vector satisfying the conditions \eqref{condz2} is
\be\label{vhm51}
v = \frac1{4}(1,1,1,1,1,1,1,1;1,1,1,1)\ . \vs{-5}
\ee
Since $v \notin I^*$, there are no massless vector multiplets in the twisted sector.

We next look at the untwisted sector. We already know that there are 8 massless vectors arising from oscillator modes.
To look for additional states we write vectors $P\in\Gamma$ as $(P_N,P_{IL},P_{IR})$, where $P_{IL}$ and $P_{IR}$ are 
respectively the left and right components of $P_I$. For massless states it must be $P_R=P_{IR}=0$ and
$P_L^2=P_N^2 + P^2_{IL}=2$.
If $P_N=0$, then $P_I \in I$ and there could be massless states with $P_I=(n_1, \cdots, n_8;0,0,0,0)$, $n_1^2 + \cdots + n_8^2=1$ 
and $n_i\in\bz$. They are all projected out because all have $e^{2i\pi P_I \cdot v}=-1$. 
If $P_{IL}=0$ then $P_N \in N$ and by construction there are no solutions of $P_N^2=2$.
However, when $P_N \not=0$, there might be additional massless states because in principle it is possible to have 
$P_N^2 + P_{IL}^2=2$, where $P_N \in N^*$ and $P_{IL} \in I^*$. 
Examining the correlated classes of $N^*/N$ and $I^*/I$ shows that there are 128 solutions with 
$P_{IR}=0$, $P_N^2 + P_{IL}^2=2$ and $P_N \not= 0$. For each $(P_N,P_{IL})$ there appears the 
corresponding $(-P_N,P_{IL})$. We can then form 64 invariant combinations that 
give extra massless vector multiplets. These happen to divide into two orthogonal sets each of 32 elements. 
Besides, in each set of 32 there are 8 with $(P_N^2, P_I^2)=(\frac32,\frac12)$ and 24 with $(P_N^2, P_I^2)=(1,1)$.

Let us finally discuss the possible gauge group. Recall that there are 8 vectors from the oscillators along $I$ which in principle correspond 
to Cartan generators, i.e. the vertex operators that involve derivatives $d X_i$, where $X_i$ denote world-sheet bosons along $I$. 
Then there are 64 generators that split into two orthogonal sets. In fact, one set has components different from zero only along 
$X_1, \ldots, X_4$, while for the other the non zero components are along $X_5, \ldots, X_8$. Moreover, as mentioned above, 
in each set of 32 elements there are 8 with $P_I^2=\frac12$ and 24 with $P_I^2=1$. Interpreting these as short and long roots 
suggests that the full gauge group is $SO(9) \times SO(9)$. Indeed, appearance of $SO(9)^2$ in $6d$ is expected from the results 
of \cite{Fraiman:2021hma} and \cite{Fraiman:2022aik}. Moreover, a model with group $SO(9)^2$, with each factor realized 
at Kac-Moody level 2, was found in the fermionic formulation \cite{Chaudhuri:1995fk}. Although we have not proved that
each $SO(9)$ appears at level 2 in our construction, it might be expected since invariant states are combinations of 
$(P_N,P_{IL})$ and $(-P_N,P_{IL})$.

\subsubsection{Example 2}\label{ex2_hm5}

The invariant lattice is taken to be $I_2\!=\!E_8(2) \oplus 4A_1(-1)$. For all $P$ belonging to $E_8^*(2) \! = \! E_8(\frac12)$, 
it happens that $e^{2i \pi P^2}\!=\!1$. Thus, the characteristic vector has no components along left-moving directions.
There is a unique shift satisfying the modular invariance conditions. In the $I_2$ basis we have
\be
w=\frac{1}{2}(0^8;1^4)\ , \qquad\qquad   v=\frac{1}{4}(0^8;1^4)\ . \vs{-5}
\ee
Since $v \notin I^*$, there are no massless states in the twisted sector. In the untwisted sector there are no solutions 
with $P_N=0$, $P_{IR}=0$ and $P_{IL}^2=2$. There are instead 144 solutions with $P_{IR}=0$, $P_N^2 + P_{IL}^2=2$ 
and $P_N \not=0$.  It turns out that we can form 72 invariant combinations which all have $P_I^2=1$ and cannot be 
subdivided into smaller orthogonal sets. Thus, it appears that the group has rank 8, from the $dX_i$ Cartan generators, and 72 
roots of equal length. This suggests that the gauge group is $SU(9)$ realized at Kac-Moody level 2. According to results of \cite{Fraiman:2021hma} and  \cite{Fraiman:2022aik}, the group $SU(9)$ can appear in a $6d$ theory with 16 supercharges.
A model with $SU(9)$ at level 2 was also reported in \cite{Chaudhuri:1995fk}.

\subsection{HM63}\label{apprr_hm63}

We next consider a $\bz_6$ asymmetric orbifold in 6 dimensions. The normal lattice $N$ has rank $s=18$ and the invariant lattice 
$I$ has signature $(2,4)$. The automorphism has an eigenvalue distribution $t=\tfrac1{6}(1,1,1,2,2,2,3,3,3)$, 
i.e. $n_1=n_2=n_3=3$. The corresponding frame shape is $2^3 6^3$ \cite{HL}. The zero point energies are
\be\label{hm63Es}
E_1=\frac{11}{12}\ ,\qquad\qquad E_2=\frac{2}{3}\ ,\qquad\qquad E_3=\frac{3}{4}\ .
\ee
The normal lattice $N$ has discriminant group $N^*/N\cong\bz_2^3\times\bz_6^3$, and there are vectors in $\bz_2$ classes of 
$N^*$ that have half-integer length squared. Therefore, the lattice $I$ must have $I^*/I\cong\bz_2^3\times\bz_6^3$ and must 
contain $\bz_2$ classes with half-integer length squared, in order to correlate with $N$ to form $\Gamma(20,4)$. 
We found that it is possible to correlate $N$ with
\vs{-5}\be\label{I63}
I=A_1\oplus A_1(3)\oplus2A_1(-1)\oplus2A_1(-3) \, , \vs{-5}
\ee
which will be considered in our analysis below. We also checked that
another possible choice\footnote{Note that there are also lattices 
with signature $(2,4)$ and $I^*/I\cong\bz_2^3\times\bz_6^3$, such that $P^2\in\bz$ for all $P\in\bz_2$ classes of $I^*$. 
Examples of this type are $U(2)\oplus U(6)\oplus A_2(-2)$ and also $A_2(2)\oplus2A_2(-2)$ which was proposed in
\cite[Table 1]{Persson:2015jka}. However, since there are no vectors with half-integer length squared in their $\bz_2$ classes, 
it is not possible to correlate these lattices with $N$ to form $\Gamma(20,4)$.}
is $I'=A_1\oplus A_1(3)\oplus2A_2(-2)$.
The generic invariant lattice has 8 moduli. For instance, in the above $I$ we can easily boost $A_1\oplus A_1(-1)$ to a 
generic $(1,1)$ lattice with one modulus.     

The invariant lattices $I$, $I_2$ and $I_3$ have
\vs{-5}\be\label{I*63}
I^*/I\cong \bz_2^3\times\bz_{6}^3 \cong \bz_2^6\times\bz_{3}^3\ ,\qquad\qquad 
I_2^*/I_2\cong\bz_{3}^6\ ,\qquad\qquad I_{3}^*/I_{3}\cong\bz_2^{12}\ .
\ee
Hence, $\ell_2=6$, $\ell_2^N=6$, $\ell_{3}=3$ and $\ell_{3}^N=3$. The degeneracy factors are
\be\label{F63}
\cF_1=1\ ,\qquad\qquad \cF_2=\cF_2'=1\ ,\qquad\qquad \cF_{3}=\cF_{3}'=1\ .\vs{-5}
\ee
The values of $C_2$ and $C_{3}(1)$ are
\be\label{C3C2hm63}
C_2=2^3\ ,\qquad\qquad\qquad C_{11}(1)=3i\sqrt{3}\ ,\vs{-5}
\ee
i.e. $\theta_2=0$ and $\theta_{3}^1=\frac12$. Consistency conditions \eqref{opinttot} are all satisfied. 

The partition function of the $T^4/\bz_6$ asymmetric orbifold in this example can be obtained from
the general results in section \ref{section_Z}. We just need to set $m=3$ and take into account that
$I$, $I_2$ and $I_3$, as well as the shifts $v$ and $w$, have left- and right-moving directions. 

The characteristic vector $w$ is determined as explained in appendix \ref{app_w} and is displayed below.
On the other hand, the shift vector $v$ must satisfy the modular invariance constraints 
$6v+w\in I$ and $4mE_1+2mv^2\in2\bz$, as well as the condition $2v \notin I^*$,  to avoid massless states in the twisted 
sectors. We find that up to equivalences there is a unique solution for $v$. The concrete results for the shifts are
\be\label{hm63shifts}
w=\frac12(1,1;1,1,1,1)\ ,\qquad\qquad\qquad\qquad 
v=\frac1{12}(1,1;3,1,1,1)\ ,
\ee
written in the basis of $I$. 

With the above $v$ there are no massless vectors in the twisted sectors.
In the untwisted sector, $m_R^2=0$ requires $P_R=0$, $N_R=0$ and $r^2=1$, cf. eq. \eqref{massunt}.
Solutions to $m_L^2=0$ which are of the type 
$P_L=0$ and $N_L=1$ give 2 vector multiplets from the two oscillators along the left-moving directions of $I$.
For solutions of the type $P_N^2+P_{IL}^2=2$ and $N_L=0$, there are no solutions with 
$P_{IL}=0$, there are 2 solutions with $P_N=0$, i.e. with $P_I \in I$ but they are projected out.
There are 22 solutions with $P_N\ne0$ and $P_{IL} \ne 0$ which come from correlated classes. 
They are precisely of the general form described in cases 1,2 and 3 in subsection \ref{subsection_op_un}. 
We find that the 4 solutions of type 1 and the 6 of type 2 are projected out. Only the 12 solutions of type 3 
survive the orbifold projection and give rise to 2 invariant states. Altogether there are 4 vector multiplets
pointing to gauge group $SU(2) \times U(1)$. At generic points of moduli space the group is just $U(1)^2$.

\small\baselineskip=.87\baselineskip
\let\bbb\bibitem\def\bibitem{\itemsep1.5pt\bbb}

\bibliographystyle{utphys}
\bibliography{Leech}

\end{document}